\documentclass[acmtog,authorversion]{acmart}

\usepackage{booktabs} 
\usepackage{xcolor} 
\usepackage{colortbl} 
\usepackage{multirow} 
\usepackage{array} 
\usepackage{float} 
\usepackage{graphicx} 
\usepackage{amsmath} 
\usepackage{xspace} 

\makeatletter
\@ifundefined{@LN@col}{%
  \newcommand{\@LN@col}[1]{}%
}{}
\makeatother

\makeatletter
\DeclareRobustCommand\onedot{\futurelet\@let@token\@onedot}
\def\@onedot{\ifx\@let@token.\else.\null\fi\xspace}

\makeatother

\newcommand{\model}{Audio-Omni\xspace}
\newcommand{\editset}{AudioEdit\xspace}

\definecolor{darkgreen}{rgb}{0.0, 0.5, 0.0} 
\definecolor{linkblue}{RGB}{30, 115, 170}

\acmJournal{TOG}

\renewcommand\footnotetextcopyrightpermission[1]{}
\setcopyright{none}
\makeatletter
\def\@acmArticleSeq{}
\def\@journalName{}
\def\@acmVolume{}
\def\@acmNumber{}
\def\@acmArticle{}
\def\@acmPubDate{}
\makeatother

\begin{document}

\makeatletter
\let\@authorsaddresses\@empty
\makeatother

\title{\model: Extending Multi-modal Understanding to Versatile Audio Generation and Editing}

\author{Zeyue Tian}
\affiliation{\institution{Hong Kong University of Science and Technology}\country{Hong Kong SAR, China}}
\email{ztianad@connect.ust.hk}

\author{Binxin Yang}
\affiliation{\institution{WeChat Vision, Tencent Inc}\country{China}}

\author{Zhaoyang Liu}
\affiliation{\institution{Hong Kong University of Science and Technology}\country{Hong Kong SAR, China}}

\author{Jiexuan Zhang}
\affiliation{\institution{Peking University}\country{Beijing, China}}

\author{Ruibin Yuan}
\affiliation{\institution{Hong Kong University of Science and Technology}\country{Hong Kong SAR, China}}

\author{Hubery Yin}
\affiliation{\institution{WeChat Vision, Tencent Inc}\country{China}}

\author{Qifeng Chen}
\affiliation{\institution{Hong Kong University of Science and Technology}\country{Hong Kong SAR, China}}

\author{Chen Li}
\affiliation{\institution{WeChat Vision, Tencent Inc}\country{China}}
\authornote{Corresponding authors.}

\author{Jing Lyu}
\affiliation{\institution{WeChat Vision, Tencent Inc}\country{China}}

\author{Wei Xue}
\authornotemark[1]
\affiliation{\institution{Hong Kong University of Science and Technology}\country{Hong Kong SAR, China}}

\author{Yike Guo}
\affiliation{\institution{Hong Kong University of Science and Technology}\country{Hong Kong SAR, China}}

\begin{abstract}
  Recent progress in multimodal models has spurred rapid advances in audio understanding, generation, and editing. However, these capabilities are typically addressed by specialized models, leaving the development of a truly unified framework that can seamlessly integrate all three tasks underexplored. While some pioneering works have explored unifying audio understanding and generation, they often remain confined to specific domains.
  To address this, we introduce \model, the first end-to-end framework to unify generation and editing across general sound, music, and speech domains, with integrated multi-modal understanding capabilities.
  Our architecture synergizes a frozen Multimodal Large Language Model for high-level reasoning with a trainable Diffusion Transformer for high-fidelity synthesis.
  To overcome the critical data scarcity in audio editing, we construct \editset, a new large-scale dataset comprising over one million meticulously curated editing pairs.
  Extensive experiments demonstrate that \model achieves state-of-the-art performance across a suite of benchmarks, outperforming prior unified approaches while achieving performance on par with or superior to specialized expert models. Beyond its core capabilities, \model exhibits remarkable inherited capabilities, including knowledge-augmented reasoning generation, in-context generation, and zero-shot cross-lingual control for audio generation, highlighting a promising direction toward universal generative audio intelligence.
  The code, model, and dataset will be publicly released on \textcolor{linkblue}{\url{https://zeyuet.github.io/Audio-Omni}}.
\end{abstract}



\begin{teaserfigure}
  \includegraphics[width=\textwidth]{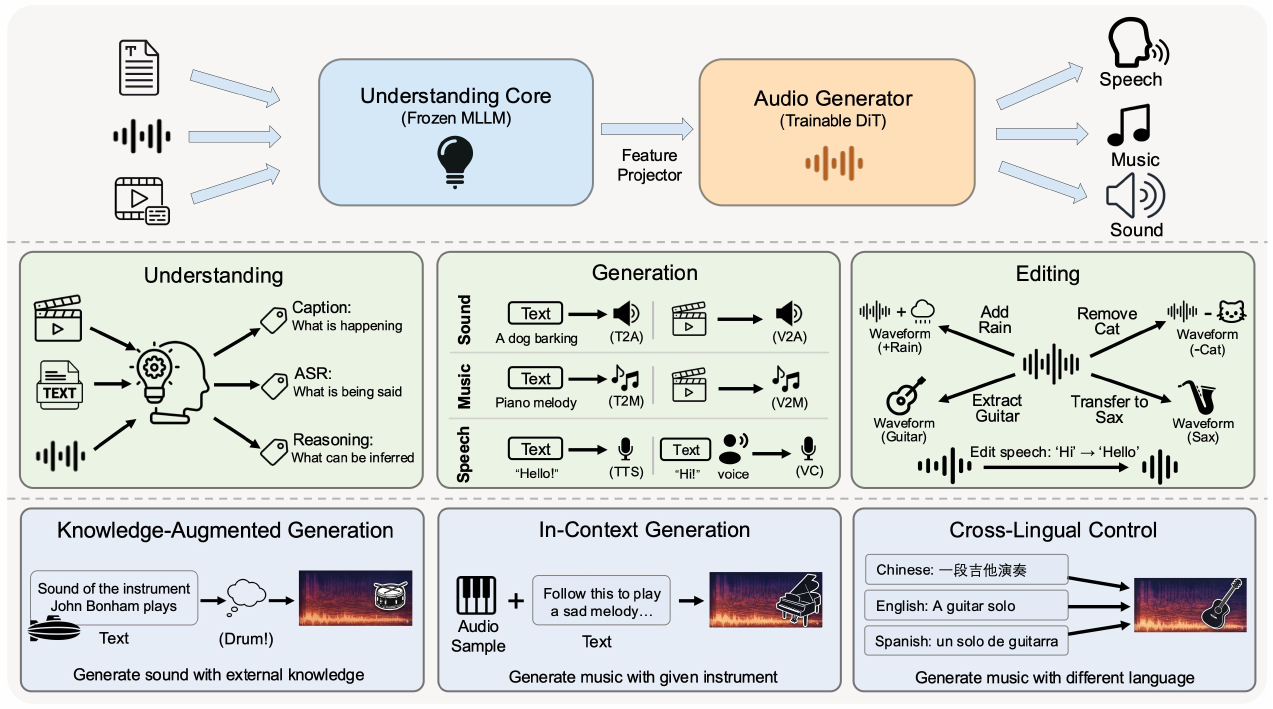} 
  \caption{\textbf{An overview of the \model framework and its capabilities.} 
  (Top) Our decoupled architecture connects a frozen MLLM for understanding with a trainable DiT for audio synthesis via a feature projector. 
  (Middle) A showcase of the model's unified capabilities across understanding, generation, and editing. 
  (Bottom) A demonstration of remarkable emergent abilities inherited from the MLLM.}
  \label{fig:teaser}
\end{teaserfigure}

\renewcommand{\shortauthors}{Tian et al.}

\makeatletter
\fancypagestyle{firstpagestyle}{%
  \fancyhf{}%
  \renewcommand{\headrulewidth}{\z@}%
  \renewcommand{\footrulewidth}{\z@}%
  \fancyhead[L]{\ACM@linecountL}%
  \fancyhead[R]{\ACM@linecountR}%
  \fancyfoot[RO,LE]{}%
}
\makeatother

\maketitle

\fancyfoot[RO,LE]{}

\section{Introduction}
\label{sec:intro}

Recent advances in multimodal learning have spurred a trend toward unified frameworks that integrate both understanding and generation within a single model, achieving significant progress in visual domains such as image~\cite{jiao2025thinkgen, ma2025janusflow, chen2025janus, pan2025transfer} and video~\cite{wei2025univideo, liu2025javisgpt} understanding and generation. 
However, the audio domain remains comparatively underexplored.

Unlike the visual modality, audio encompasses three distinct domains with significant distributional disparities: general sounds, music, and speech. While some efforts have been made to unify audio understanding and generation, they remain confined to specific domains, such as speech-centric models~\cite{an2024funaudiollm, ai2025ming} or those limited to general audio and music~\cite{tian2025ualm, liu2024mumu}, failing to cover the full audio spectrum. Other systems rely on tool-based integration~\cite{huang2024audiogpt}, which lacks end-to-end optimization. Meanwhile, existing audio editing models~\cite{manor2024zero, wang2023audit} are designed exclusively for editing and cannot be extended to understanding or generation, leaving the unification of all three capabilities an open challenge.

To address these limitations, we introduce \model, a framework that unifies audio understanding, generation, and editing across the full spectrum of audio domains.
We adopt a decoupled design: a frozen Multimodal Large Language Model (MLLM) serves as the reasoning core, while a trainable Diffusion Transformer (DiT) handles generation and editing. Keeping the MLLM frozen preserves its rich multimodal knowledge, which in turn empowers the generative module with capabilities beyond its explicit training scope. To effectively bridge the two components, we design a hybrid conditioning mechanism that disentangles inputs into two complementary streams: a \textbf{High-Level Semantic} stream, combining MLLM features and text embeddings for speech synthesis, injected via cross-attention to provide instructional guidance; and a \textbf{Low-Level Signal} stream, fusing mel-spectrogram and video sync features, concatenated with the input noise for precise temporal control. This separation is key to mastering the diverse requirements of sound, music, and speech within a single framework.

Training a unified model of this scope demands a comprehensive and diverse dataset. A critical barrier to progress in instruction-guided audio editing is the absence of any large-scale, publicly available dataset. To address this gap, we meticulously design a pipeline to construct \editset, a large-scale, high-quality dataset for this task. Created through a systematic pipeline combining real-world data mining with scalable programmatic synthesis, \editset contains over 1M rigorously curated samples covering editing tasks including addition, removal, extraction, and style transfer. Training on this dataset enables our model with its robust editing capabilities.

Extensive experiments validate the effectiveness of our unified design. \model outperforms prior unified models across understanding, generation, and editing tasks, while matching or surpassing specialized expert models on several tasks. Beyond these core results, the generative module naturally inherits capabilities from the frozen MLLM, including world knowledge for reasoning-based generation, in-context learning for audio-conditioned synthesis, and multilingual understanding for cross-lingual control.

In summary, our main contributions are as follows:
\begin{itemize}
    \item We propose \model, the first unified framework for audio understanding, generation, and editing across general sound, music, and speech. At its core is a decoupled architecture that bridges a frozen MLLM with a trainable DiT, guided by a hybrid conditioning mechanism to disentangle semantic and signal-level control.
    
    \item We introduce \editset, a large-scale, high-quality dataset with a meticulous pipeline for instruction-guided audio editing, encompassing a wide range of tasks to facilitate future research in this area.
    
    \item Extensive experiments demonstrate that \model outperforms prior unified models and achieves competitive or superior results compared to specialized expert models across a broad range of tasks.
    
    \item Furthermore, \model exhibits remarkable inherited capabilities for generation, such as knowledge-augmented generation and cross-lingual control, highlighting a path toward intelligent and versatile generative audio systems.
\end{itemize}

\section{Related Work}

\noindent\textbf{Specialized Models in the Audio Domain.}
In recent years, the field of audio processing has achieved significant advancements, particularly with the advent of large-scale pre-trained models. For audio understanding, a substantial body of work has produced powerful models capable of tackling complex tasks across all three audio domains~\cite{DBLP:conf/icml/KongGBPVC24, DBLP:conf/icml/GhoshKKSKPVMC25, chu2023qwen, chu2024qwen2, DBLP:conf/iclr/TangYSC000M024}. In contrast to the comprehensive domain coverage in understanding, generative models have typically specialized. In the general audio domain, prominent text-to-audio models like~\cite{huang2023make, ghosal2023text, majumder2024tango, evans2024stable} have demonstrated high-fidelity synthesis from textual descriptions, while other approaches have explored generation conditioned on video~\cite{xing2024seeing, cheng2025mmaudio, liu2025thinksound, liu2024tell}. Similarly, in the music domain, models like~\cite{copet2024simple, melechovsky2024mustango, chen2024musicldm, yuan2024chatmusician, yuan2025yue, he2024llms, yinghao2024foundation, deng2024composerx} excel at text-to-music generation, and dedicated research has addressed video-to-music generation~\cite{tian2025vidmuse, su2023v2meow, lin2024vmas, xie2025filmcomposer}. The speech domain has its own rich landscape of research, with powerful text-to-speech (TTS) systems~\cite{du2024cosyvoice, chen2025f5, deng2025indextts} and models that support complex transformations like voice conversion (VC)~\cite{wang2023neural, peng2024voicecraft}. While some recent models demonstrate capabilities in handling multimodal conditions or generating audio across multiple domains~\cite{tian2025audiox, wang2025audiogen, liu2025unimoe, rong2025audiogenie}.
Meanwhile, the field of audio editing remains underexplored, largely due to a scarcity of large-scale, instruction-guided datasets. Existing approaches fall into two main categories. Training-free, zero-shot methods adapt pre-trained models but often struggle with preserving non-target content and following precise instructions~\cite{manor2024zero}. Training-based methods, such as~\cite{wang2023audit, lan2025guiding, tao2025mmedit}, construct synthetic data pipelines but introduce a significant domain gap, as their strategy of mixing isolated audio segments fails to capture the acoustic complexities of real-world editing scenarios.

\begin{figure*}[t]
    \centering
    \includegraphics[width=0.9\textwidth]{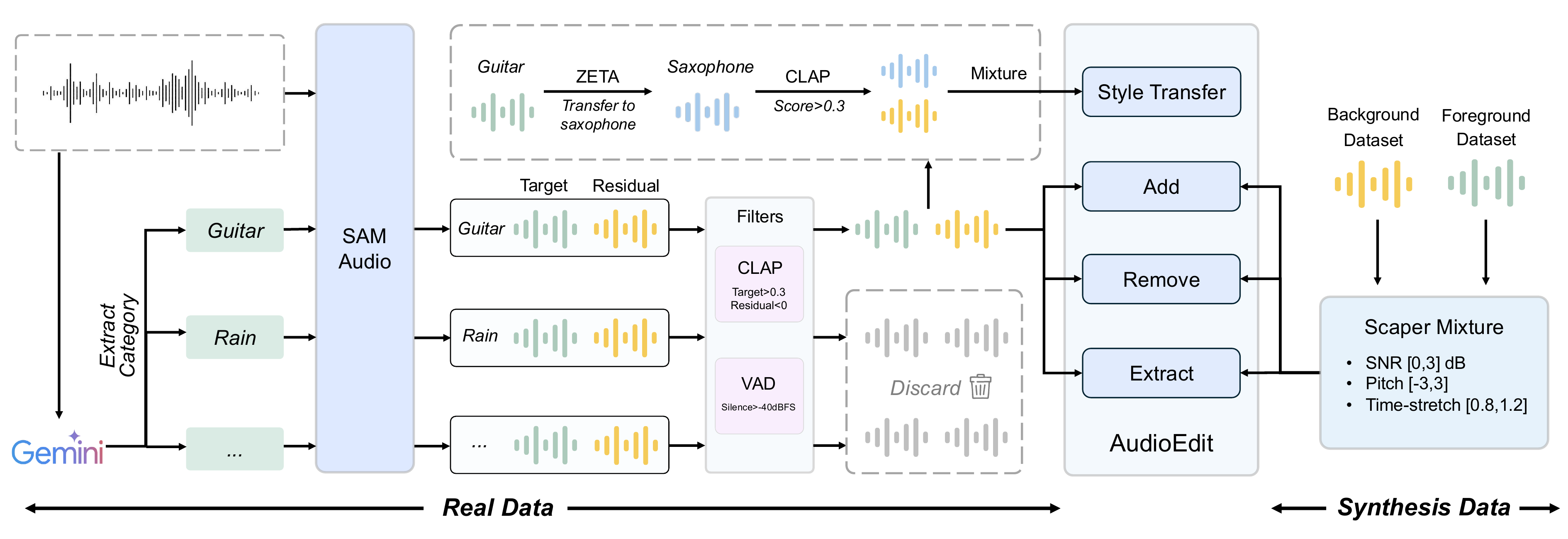}
    \caption{
        \textbf{ Overview of the hybrid pipeline for constructing our \editset dataset.} 
        The pipeline consists of two parallel branches to ensure both data authenticity and scale. The Real Data Branch (left) mines editing pairs from real-world datasets (e.g., VGGSound) by first using an MLLM (Gemini) for category identification, followed by a dedicated segmentation model (SAM-Audio) for source separation. Concurrently, the Synthesis Data Branch (right) leverages the Scaper toolkit to programmatically generate a large volume of precisely annotated editing scenarios. This hybrid strategy yields a dataset that combines the acoustic fidelity of natural audio with the large-scale diversity needed for robust model training.
    }
    \label{fig:dataset_pipeline}
\end{figure*}

\noindent\textbf{Unified Multimodal Models.}
Recently, a major trend in multimodal research has been the development of unified models that aim to handle both understanding and generation within a single framework. This unification has seen remarkable success in visual domains, with models like~\cite{pan2025transfer, jiao2025thinkgen, ma2025janusflow, chen2025janus} achieving unified image understanding and generation, and others extending these capabilities to the video domain~\cite{wei2025univideo, liu2025javisgpt}. Some frameworks even support generation across different modalities, such as text, images, audio, and video~\cite{zhan2024anygpt, wu2024next, lu2024unified}.
In the audio field, several pioneering works~\cite{DBLP:conf/aaai/HuangLYSCYWHHLR24, an2024funaudiollm, liu2024mumu, ai2025ming} have also moved towards this unified goal. However, they exhibit critical limitations. Some approaches rely on orchestrating separate expert models via tool invocation~\cite{DBLP:conf/aaai/HuangLYSCYWHHLR24}, which lacks the benefits of end-to-end optimization. Other models, while more integrated, typically focus on a limited subset of audio domains, such as speech-only~\cite{an2024funaudiollm, ai2025ming} or music-only generation~\cite{liu2024mumu}, failing to provide a truly universal solution. To address these shortcomings, we propose \model, the first framework to unify understanding, generation, and editing across all general sound, music, and speech domains. Our approach provides a cohesive, end-to-end solution that overcomes the fragmentation of specialized models and the domain limitations of existing unified efforts.

\section{Audio Editing Dataset Construction}
\label{sec:dataset}

\begin{table}[h!]
\centering
\small
\caption{
    Statistics of our proposed audio editing dataset \editset.
}
\label{tab:dataset_stats}
\resizebox{\columnwidth}{!}{
\begin{tabular}{ll rr} 
\toprule
\textbf{Task Type} & \textbf{Data Source} & \textbf{Train Samples} & \textbf{Test Samples} \\
\midrule
\multirow{2}{*}{\textbf{Add}} & Real Data & 50K & 500 \\
 & Synthesis Data & 150K & - \\
\midrule
\multirow{2}{*}{\textbf{Remove}} & Real Data & 50K & 500 \\
 & Synthesis Data & 150K & - \\
\midrule
\multirow{2}{*}{\textbf{Extract}} & Real Data & 50K & 500 \\
 & Synthesis Data & 150K & - \\
\midrule
\textbf{Style Transfer} & Real Data & 500K & 500 \\
\midrule
\textbf{Total} & & \textbf{1,100K} & \textbf{2,000} \\
\bottomrule
\end{tabular}
}
\end{table}

An obstacle to advancing instruction-guided audio editing is the scarcity of large-scale paired datasets. While pioneering works like~\cite{wang2023audit, lan2025guiding, tao2025mmedit} have proposed synthetic pipelines, their reliance on mixing isolated audio segments introduces a significant domain gap from real-world audio, where sounds are integrated. Furthermore, these methods often falter on complex tasks like audio style transfer, struggling to disentangle acoustic style from core content attributes such as temporal structure and pitch. To address these limitations, we introduce \editset, a large-scale dataset constructed via a novel hybrid pipeline that integrates real-world data mining with scalable synthesis.

Our pipeline, illustrated in Figure~\ref{fig:dataset_pipeline}, features two complementary branches to generate a dataset with both real-world acoustic fidelity and large-scale diversity.
\textbf{The Real Data Branch} mines authentic editing pairs from real-world datasets VGGSound~\cite{chen2020vggsound}.
\textit{First}, we employ a powerful MLLM (Gemini 2.5 Pro) to identify the primary sound-emitting object categories within each source audio clip.
\textit{Then}, we deploy SAM-Audio~\cite{shi2025sam}, a state-of-the-art audio segmentation model, to perform source separation based on the identified categories. This step disentangles the source audio into a target track, containing the audio of the specified object, and a corresponding residual track.
\textit{Subsequently}, all separated tracks undergo rigorous multi-stage filtering to ensure quality. Starting from over 540K category-labeled samples, we apply VAD\footnote{\url{https://github.com/jiaaro/pydub}} filtering (retaining $\sim$347K pairs) followed by CLAP~\cite{elizalde2023clap}-based semantic alignment (retaining $\sim$50K pairs, approximately 9.2\% overall retention). Human validation on a subset achieves approximately 83\% agreement, confirming pipeline quality.
This process yields high-quality source-target pairs for \texttt{add}, \texttt{remove}, and \texttt{extract} tasks. 
For \texttt{style transfer}, we expand the filtered targets by prompting Gemini to generate semantically related but different keywords (prompt template in Appendix~\ref{sec:appendix_dataset_details}), then apply CLAP filtering again, yielding approximately 500K pairs. We use ZETA~\cite{manor2024zero} to transform each target to the new style while preserving temporal structure and pitch, then mix the transformed audio back with the residual track to form the final edited output.
Concurrently, our \textbf{Synthesis Data Branch} ensures scale and diversity by programmatically generating soundscapes with the Scaper toolkit~\cite{salamon2017scaper}. This is achieved by randomly mixing foreground events from ESC-50~\cite{piczak2015esc} into 10-second backgrounds from AudioCaps~\cite{kim2019audiocaps}, applying randomized parameters including onset time, SNR (0-3 dB), pitch shifts (-3 to +3 semitones), and time-stretch factors (0.8-1.2). This automated process efficiently yields a large volume of precisely annotated data for our \texttt{add}, \texttt{remove}, and \texttt{extract} tasks.

Through this meticulously designed hybrid pipeline, we construct \editset, a large-scale, high-quality dataset for instruction-guided audio editing. 
In total, \editset comprises over 1M samples covering \texttt{add}, \texttt{remove}, \texttt{extract}, and \texttt{style transfer} audio editing tasks, with detailed statistics presented in Table~\ref{tab:dataset_stats}.
\section{Method}
\label{sec:method}

In this section, we describe the architecture of \model and its training strategies.

\subsection{Preliminary}
\label{sec:preliminary}

\noindent\textbf{Rectified Flow.}
Our generative backbone is built upon the framework of Rectified Flow~\cite{liu2022flow}, a powerful class of generative models. 
Unlike traditional diffusion models that often follow stochastic paths, Rectified Flow simplifies the generation process by modeling a straight-line trajectory between noise and data. 
This trajectory connects a random noise sample $\mathbf{x}_1 \sim \mathcal{N}(\mathbf{0}, \mathbf{I})$ to a data sample $\mathbf{x}_0$ via a simple ordinary differential equation (ODE) with a constant velocity field $\mathbf{v} = \mathbf{x}_1 - \mathbf{x}_0$. 
The ODE is defined for a time variable $t \in [0, 1]$ as:
\begin{equation}
    \frac{d\mathbf{x}_t}{dt} = \mathbf{v}.
    \label{eq:rf_ode}
\end{equation}
The solution at any time $t$ along this linear path is given by the interpolation:
\begin{equation}
    \mathbf{x}_t = (1-t)\mathbf{x}_0 + t\mathbf{x}_1.
    \label{eq:rf_path}
\end{equation}
A neural network, denoted $v_{\theta}(\mathbf{x}_t, t, \mathbf{c})$, is trained to predict this velocity field $\mathbf{v}$ conditioned on the noisy state $\mathbf{x}_t$, time $t$, and a set of conditioning signals $\mathbf{c}$. 
During inference, generation starts from a random noise sample $\mathbf{x}_1$ at $t=1$, and the ODE in Equation~\ref{eq:rf_ode} is solved backwards to $t=0$ using a numerical solver, guided by the predictions of $v_{\theta}$. 
The training objective for $v_{\theta}$ is detailed in Section~\ref{sec:training}.

\begin{figure*}[t]
    \centering
    \includegraphics[width=0.9\textwidth]{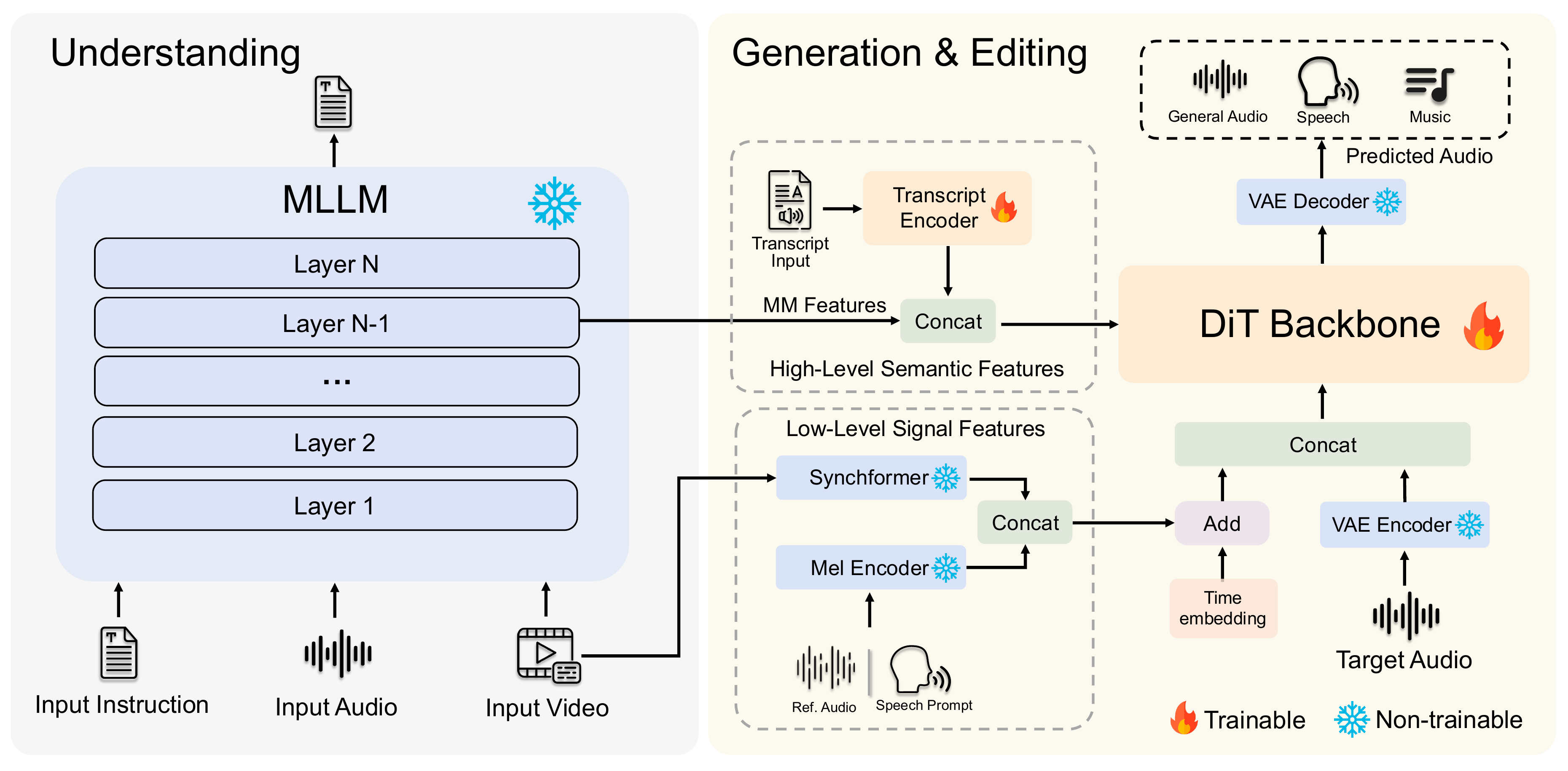}
    \caption{
        \textbf{The \model Framework.}
        Our framework utilizes a decoupled design with two distinct conditioning streams to guide a trainable DiT backbone. The High-Level Semantic Features stream provides global, instructional guidance. It is formed by concatenating features from a frozen MLLM (MM Features) with character-level embeddings from a trainable Transcript Encoder. The Low-Level Signal Features stream offers precise, temporal guidance for editing and synchronization. It combines features from Synchformer and Mel Encoder. These two streams are injected into the DiT via different mechanisms: the high-level stream as context for cross-attention, and the low-level stream concatenated with the input noise and time embedding.
    }
    \label{fig:method-omni}
\end{figure*}

\subsection{Model Architecture}
\label{sec:model_architecture}

As illustrated in Figure~\ref{fig:method-omni}, our \model framework is built upon a decoupled architecture comprising a frozen MLLM for understanding and a trainable DiT-based backbone for versatile audio generation and editing.

\noindent\textbf{Understanding Module.}
Our understanding module is a pre-trained and frozen MLLM, which serves as the primary reasoning and understanding core of our framework. It processes a diverse set of inputs, including a textual instruction ($\mathbf{T}_{\text{in}}$), an audio waveform ($\mathbf{A}_{\text{in}}$), and a video ($\mathbf{V}_{\text{in}}$), after they are tokenized by their respective encoders. 
The MLLM fulfills a dual role within our framework. For \textit{understanding tasks}, it directly generates textual responses based on the multimodal inputs. For \textit{generative tasks}, its crucial function is to produce a powerful conditioning signal for the generative module. To this end, we extract the hidden states from its penultimate layer (i.e., the second-to-last hidden state), as we empirically find that this provides a more generalizable and richer semantic representation for downstream generative tasks compared to the final layer (see Sec.~\ref{sec:ablation_studies}). We denote this multimodal feature as $\mathbf{F}_{\text{mm}} \in \mathbb{R}^{L_{\text{mm}} \times D_{\text{mm}}}$.

\noindent\textbf{Generative Module.}
The generative module consists of a DiT backbone, trained with a Rectified Flow objective, which synthesizes the final audio waveform. This module is conditioned on two distinct streams of features, allowing it to handle a wide array of tasks.

The first stream, which we term \textit{High-Level Semantic Features}, provides the primary instructional signal for the synthesis process. This stream is formed by concatenating the multimodal features $\mathbf{F}_{\text{mm}}$ from the MLLM with a transcript-derived feature $\mathbf{F}_{\text{trans}}$ for speech-related tasks:
\begin{equation}
    \mathbf{c}_{\text{high}} = \text{Concat}(\mathbf{F}_{\text{mm}}, \mathbf{F}_{\text{trans}}).
\end{equation}
The feature $\mathbf{F}_{\text{trans}}$ is produced by a Transcript Encoder, which performs a character-level encoding of the input transcript $\mathbf{T}_{\text{trans}}$ (for TTS/VC tasks). Specifically, it first converts the text into a sequence of character embeddings, which are then processed by a ConvNeXtV2-based architecture.

The second stream, termed \textit{Low-Level Signal Features}, provides concrete, temporally-aligned references crucial for editing and synchronization tasks. This stream is formed by concatenating features from a reference audio and a video sync signal:
\begin{equation}
    \mathbf{c}_{\text{low}} = \text{Concat}(\mathbf{F}_{\text{sync}}, \mathbf{F}_{\text{mel}}).
\end{equation}
The mel-spectrogram feature, $\mathbf{F}_{\text{mel}}$, is extracted by a Mel Encoder from either a reference audio $\mathbf{A}_{\text{ref}}$ (for editing) or a speech prompt $\mathbf{S}_{\text{prompt}}$ (for voice conversion). The synchronization feature, $\mathbf{F}_{\text{sync}}$, is extracted from the input video $\mathbf{V}_{\text{in}}$ by a pre-trained Synchformer~\cite{iashin2024synchformer} model.

These two conditioning streams are injected into the DiT backbone through distinct mechanisms tailored to their properties. The \textit{High-Level Semantic Features} ($\mathbf{c}_{\text{high}}$) are injected as context via cross-attention, allowing the model to flexibly query abstract instructions at each step of the generation process. In contrast, the \textit{Low-Level Signal Features} ($\mathbf{c}_{\text{low}}$) are first fused with a time embedding via element-wise addition, and this combined signal is then concatenated with the VAE-encoded noisy audio latent ($\mathbf{x}_t$) to form the main input to the DiT. This concatenation provides strong, frame-by-frame guidance, ensuring precise alignment for editing and synchronization tasks. The DiT backbone then processes this composite input to predict the velocity field, and its final denoised latent output is passed to a VAE Decoder to reconstruct the audio waveform.

\subsection{Training Objective}
\label{sec:training}

We train our \model end-to-end using a single, unified loss based on the Rectified Flow objective. 
For each training sample, we aim to train the network $v_{\theta}$ to predict the constant velocity $\mathbf{v} = \mathbf{x}_1 - \mathbf{x}_0$, where $\mathbf{x}_0$ is the VAE-encoded latent of the target audio and $\mathbf{x}_1$ is a random noise sample from $\mathcal{N}(\mathbf{0}, \mathbf{I})$.
Following the standard RF training procedure, we first sample a random timestep $t$ from a uniform distribution $\mathcal{U}(0, 1)$. This timestep is used to create an interpolated latent state $\mathbf{x}_t = (1-t)\mathbf{x}_0 + t\mathbf{x}_1$.
The training objective is then to minimize the mean squared error between the network's predicted velocity and the ground-truth velocity $\mathbf{v}$. The loss function, $\mathcal{L}$, is defined as:
\begin{equation}
    \mathcal{L} = \mathbb{E}_{t \sim \mathcal{U}(0,1), \mathbf{x}_0, \mathbf{x}_1, \mathbf{c}} \left[ || v_{\theta}(\mathbf{x}_t, t, \mathbf{c}) - (\mathbf{x}_1 - \mathbf{x}_0) ||^2 \right]
    \label{eq:rf_loss}
\end{equation}

where $\mathbf{c}$ represents the full set of conditioning signals available for the given training sample.

\section{Experiments}
\label{sec:experiment}

\subsection{Implementation Details}
\label{sec:implementation_details}

\begin{table}[t]
    \centering
    \footnotesize
    \caption{
        Quantitative results on multimodal understanding benchmarks.
    }
    \renewcommand{\arraystretch}{0.85}
    \setlength{\tabcolsep}{8pt}
    \begin{tabular}{l cc}
    \toprule
    Methods & MMSU$\uparrow$ & MMAU$\uparrow$ \\
    \midrule
    \multicolumn{3}{c}{\textit{Specialized Models}} \\
    Audio Flamingo3~\cite{goel2025audio} & \textbf{61.40} & \textbf{72.42} \\
    Qwen2-Audio-Instruct~\cite{chu2024qwen2} & 53.27 & 57.40 \\
    Qwen2.5-Omni-3B~\cite{xu2025qwen2} & \underline{56.83} & 63.30 \\    
    \cmidrule(lr){1-3}
    \multicolumn{3}{c}{\textit{Unified Models}} \\
    Ming-Omni~\cite{ai2025ming} & 47.53 & \underline{70.80} \\
    Unified-IO2~\cite{lu2024unified} & 30.74 & 3.20 \\
    MuMuLLaMA~\cite{liu2024mumu} & 6.32 & 12.87 \\
    \cellcolor{cyan!7} \model & \cellcolor{cyan!7} \underline{56.83} & \cellcolor{cyan!7} 63.30 \\
    \bottomrule
    \end{tabular}
    \label{tab:understanding}
\end{table}

\begin{table}[t]
    \centering
    \normalsize
    \caption{
        Quantitative results on multimodal generation benchmarks.
    }
    \renewcommand{\arraystretch}{1.1}
    \resizebox{\columnwidth}{!}{
    \setlength{\tabcolsep}{3pt}
    \begin{tabular}{l ccccc}
    \toprule
    Methods & T2A\_FAD$\downarrow$ & T2M\_FAD$\downarrow$ & V2A\_FAD$\downarrow$ & V2M\_FAD$\downarrow$ & TTS\_WER$\downarrow$ \\
    \midrule
    \multicolumn{6}{c}{\textit{Specialized Models}} \\
    Tango2~\cite{majumder2024tango} & 3.20 & - & - & - & - \\
    MMAudio~\cite{cheng2025mmaudio} & 4.71 & - & 2.04 & - & - \\
    Stable-Audio-Open~\cite{evans2024stable} & 3.15 & 3.23 & - & - & - \\
    MusicGen~\cite{copet2024simple} & - & 3.94 & - & - & - \\
    VATT~\cite{liu2024tell} & - & - & 2.55 & - & - \\
    AudioX~\cite{tian2025audiox} & \textbf{1.86} & \textbf{1.53} & \textbf{1.13} & \underline{2.12} & - \\
    VidMuse~\cite{tian2025vidmuse} & - & - & - & 2.46 & - \\
    F5-TTS~\cite{chen2025f5} & - & - & - & - & \underline{1.83} \\
    MaskGCT~\cite{wang2024maskgct} & - & - & - & - & 2.62 \\
    CosyVoice3~\cite{du2025cosyvoice} & - & - & - & - & 2.46 \\
    \cmidrule(lr){1-6}
    \multicolumn{6}{c}{\textit{Unified Models}} \\
    Ming-Omni~\cite{ai2025ming} & - & - & - & - & 4.31 \\
    Unified-IO2~\cite{lu2024unified} & 7.81 & 3.17 & - & - & 21.63 \\
    MuMuLLaMA~\cite{liu2024mumu} & - & 5.89 & - & 52.25 & - \\
    \cellcolor{cyan!7} \model & \cellcolor{cyan!7} \textbf{1.86} & \cellcolor{cyan!7} \underline{1.94} & \cellcolor{cyan!7} \underline{1.71} & \cellcolor{cyan!7} \textbf{1.58} & \cellcolor{cyan!7} \textbf{1.77} \\
    \bottomrule
    \end{tabular}
    }
    \label{tab:generation}
\end{table}

\begin{table}[t]
    \centering
    \small
    \caption{
        Quantitative results on audio editing benchmarks.
    }
    \renewcommand{\arraystretch}{0.85}
    \setlength{\tabcolsep}{6pt}
    \begin{tabular}{l ccc}
    \toprule
    Methods & AE\_FAD$\downarrow$ & AE\_LSD$\downarrow$ & CLAP$\uparrow$ \\
    \midrule
    \multicolumn{4}{c}{\textit{Specialized Models}} \\
    ZETA~\cite{manor2024zero} & \underline{3.81} & \underline{3.80} & \underline{0.30} \\
    SDEdit~\cite{meng2021sdedit} & 3.51 & 4.40 & 0.22 \\
    MMEDIT~\cite{tao2025mmedit} & 3.95 & 4.05 & 0.15 \\
    \cmidrule(lr){1-4}
    \multicolumn{4}{c}{\textit{Unified Models}} \\
    \cellcolor{cyan!7} \model & \cellcolor{cyan!7} \textbf{3.27} & \cellcolor{cyan!7} \textbf{2.27} & \cellcolor{cyan!7} \textbf{0.32} \\
    \bottomrule
    \end{tabular}
    \label{tab:editing}
\end{table}

\noindent\textbf{Model Architecture.}
For the MLLM, we use a pre-trained Qwen2.5-Omni-3B model~\cite{xu2025qwen2}, which processes input videos at 5 fps and audio at 16 kHz. We extract features from its penultimate layer as the primary semantic condition.
The DiT backbone is a transformer architecture with a depth of 36 blocks, a hidden dimension of 2048, and 32 attention heads. 
The entire model has approximately 7.9B total parameters, of which 3.05B (the DiT and conditioners) are trainable.
Our conditioning modules include a Synchformer~\cite{iashin2024synchformer} that extracts synchronization features from videos at 25 fps; a Mel Encoder that computes 100-dimensional mel-spectrograms (44.1 kHz, FFT size 1024, hop 256); and a Transcript Encoder composed of four ConvNeXtV2 blocks for character-level text encoding. 
The final audio is encoded and decoded by a pre-trained VAE adopted from~\cite{evans2024stable}.

\noindent\textbf{Datasets.}
The training data sources and their scale are summarized in Appendix Table~\ref{tab:training_datasets}.

\noindent\textbf{Training Details.}
We train the DiT backbone and all learnable conditioning modules for approximately 80k steps on a total batch size of 5120. 
We use the AdamW optimizer with a learning rate of 5e-5, with $\beta_1=0.9$, $\beta_2=0.999$, and a weight decay of 1e-3. 
The entire model is trained end-to-end using the Rectified Flow objective (Equation~\ref{eq:rf_loss}).
To enhance voice conversion and speech editing capabilities, we employ a masking strategy on the \textit{speech prompt} during training following~\cite{chen2025f5}. Specifically, we randomly mask 20\% to 75\% of the prompt's mel-spectrogram, forcing the model to infer the global speaker timbre from a partial acoustic signal while reconstructing the full utterance using the complete transcript. This scheme is crucial for developing the model's robust zero-shot voice cloning and content editing abilities.

\noindent\textbf{Inference Details.}
At inference time, we use a numerical ODE solver with 100 steps to generate the audio latent from a random noise sample. 
For conditional generation, we employ classifier-free guidance with a guidance scale of 6.0.

\subsection{Evaluation Metrics}
\label{sec:evaluation_metrics}

To evaluate \model, we employ a comprehensive suite of metrics for its understanding, generation, and editing capabilities. For understanding, we report scores on the MMSU~\cite{wang2025mmsu} and MMAU~\cite{sakshi2024mmau} benchmarks to assess multi-task reasoning. For generation, we measure distributional similarity and quality using KL divergence, Inception Score (IS), Fréchet Audio Distance (FAD), and Fréchet Distance (FD) on PANNs~\cite{kong2020panns} embeddings. Speech synthesis performance is specifically evaluated using Word Error Rate (WER). For editing, we use Log-Spectral Distance (LSD) to measure fidelity and content preservation, and FAD to assess perceptual quality.

\subsection{Main Results}
\label{sec:main_results}

In this section, we present quantitative and qualitative results for \model, evaluating its performance on understanding, generation, and editing tasks against a wide range of specialized and unified models.

\subsubsection{Overall Performance}
As shown in Table~\ref{tab:understanding}, Table~\ref{tab:generation}, and Table~\ref{tab:editing}, \model demonstrates strong and comprehensive capabilities across understanding, generation, and editing tasks.

\noindent\textbf{Understanding.}
Benefiting from our decoupled architecture, \model inherits a strong understanding ability from the frozen MLLM core, whose audio encoder has been pre-trained on a large-scale audio-related data, enabling strong performance on multi-domain audio tasks~\cite{xu2025qwen2}. 
We evaluate on MMSU (covering 47 spoken-language tasks) and MMAU (evaluating 27 reasoning skills across sound, music, and speech). As shown in Table~\ref{tab:understanding}, \model achieves strong understanding performance, outperforming most unified models and approaching the level of dedicated understanding specialists.

\noindent\textbf{Generation and Editing.}
For generation tasks, \model exhibits state-of-the-art or highly competitive performance. 
We report the FAD on standard test sets: AudioCaps for T2A, Musicaps for T2M, VGGSound for V2A, and V2M-bench for V2M. For speech synthesis, we evaluate WER on the Seed-TTS~\cite{anastassiou2024seed} en benchmark. 
In generation tasks, our model demonstrates strong overall performance, consistently and significantly outperforming all other unified models. Notably, it further surpasses specialized expert models in T2M and TTS.
For editing, we report the average FAD and LSD across four tasks (\texttt{add}, \texttt{remove}, \texttt{extract}, \texttt{style transfer}) on our proposed \editset test set. We additionally report the CLAP score (averaged over \texttt{add}, \texttt{extract}, and \texttt{style transfer}) to evaluate instruction adherence. As shown in Table~\ref{tab:editing}, \model achieves the best performance on all metrics, with detailed per-task results in Appendix Table~\ref{tab:appendix_editing_tasks}.

In summary, Table~\ref{tab:understanding}, Table~\ref{tab:generation}, and Table~\ref{tab:editing} validate the effectiveness of our unified framework: \model inherits strong understanding capabilities from the frozen MLLM while achieving state-of-the-art or highly competitive results across generation and editing tasks, demonstrating that a single unified model can serve as a strong generalist across the full spectrum of audio tasks.

\subsubsection{Zero-shot Cross-lingual Text-to-Audio Generation}

A remarkable ability of our framework is its zero-shot cross-lingual generation, inherited from the frozen MLLM's multilingual understanding. As shown in Appendix Table~\ref{tab:crosslingual_t2a}, \model maintains strong performance across multiple languages (CN, ES, DE, FR, JP) despite being trained on English, with quality comparable to English-only specialist models.

\subsubsection{Inherited Abilities and Zero-Shot Capabilities}

\begin{figure}[htb]
    \centering
    \includegraphics[width=\columnwidth]{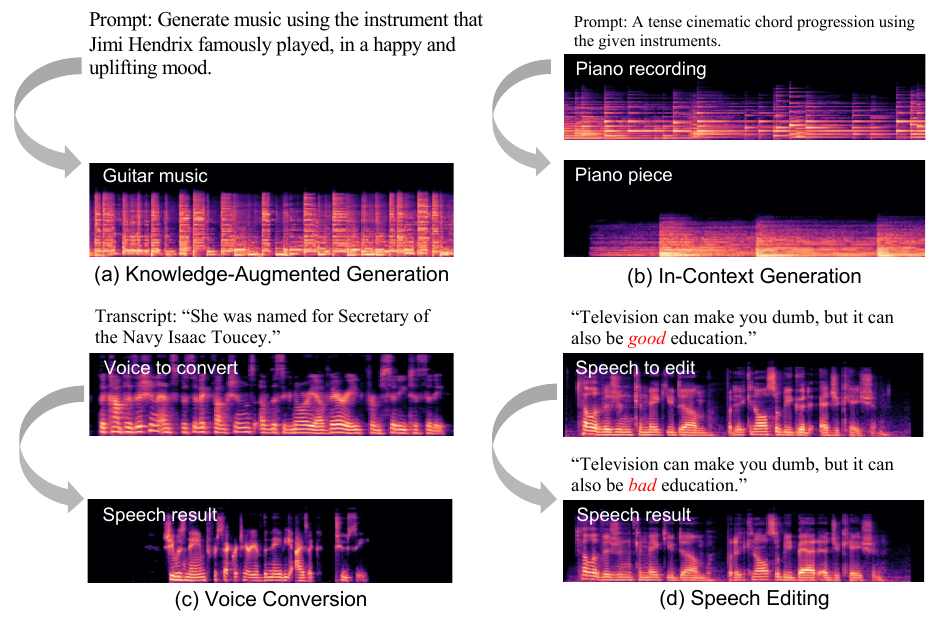}
    \caption{
        \textbf{Qualitative showcase of \model's capabilities}, including (a) knowledge-augmented generation, (b) in-context generation, (c) zero-shot voice conversion, and (d) zero-shot speech editing.
    }
    \label{fig:qualitative_examples}
\end{figure}

We further highlight several representative capabilities of \model. The first two are emergent abilities inherited from the frozen MLLM's world knowledge and in-context reasoning, while the latter two are enabled by our masking-based training strategy. We qualitatively showcase these in Figure~\ref{fig:qualitative_examples}.

\noindent\textbf{Knowledge-Augmented Generation.}
\model successfully handles knowledge-intensive prompts that require external world knowledge. For instance, when prompted with ``\textit{Generate music using the instrument Jimi Hendrix played in a happy mood.}'' it correctly infers the instrument is an electric guitar and synthesizes a corresponding melody, a task beyond standard text-to-audio models.

\noindent\textbf{In-Context Generation.}
Our model demonstrates strong in-context learning. Given a piano recording and the instruction ``\textit{Generate a dramatic cinematic chord progression that builds tension},'' it extracts the piano's timbre and applies it to the newly synthesized piece.

\noindent\textbf{Zero-Shot Voice Conversion and Speech Editing.}
During speech training, we randomly mask either speaker identity or speech content, forcing the model to infer the missing component from context. This naturally enables zero-shot voice conversion and speech editing at inference time without task-specific supervision.

These capabilities demonstrate that \model goes beyond a standard multi-task model, exhibiting inherited intelligence and flexible zero-shot control across diverse audio generation scenarios.

\subsection{Ablation Studies}
\label{sec:ablation_studies}

To validate our design choices, we conduct a series of ablation studies in this section, more ablation studies are included in the Appendix~\ref{sec:appendix_ablation_studies}.

\vspace{1mm}
\noindent\textbf{Ablation on Dataset Composition.}
To assess the impact of our data sources, we evaluate three training configurations: using only synthetic data, only real-world data, or a mix of both. As shown in Table~\ref{tab:ablation_dataset}, the mixed-data approach yields the best overall performance. This suggests a synergy where synthetic data provides broad coverage of diverse editing operations, while real-world data contributes crucial acoustic realism and fidelity. Notably, training on synthetic data alone is insufficient for achieving robust generalization to the acoustic complexities of real-world audio.
\begin{table}[t]
    \centering
    \small
    \caption{
        Ablation study on dataset composition for audio editing training.
    }
    \label{tab:ablation_dataset}
    \renewcommand{\arraystretch}{0.9}
    \setlength{\tabcolsep}{4pt}
    \begin{tabular}{l ccccc}
    \toprule
    \textbf{Setting} & \textbf{KL$\downarrow$} & \textbf{IS$\uparrow$} & \textbf{FD$\downarrow$} & \textbf{FAD$\downarrow$} & \textbf{LSD$\downarrow$} \\
    \midrule
    Only Real & \textbf{1.27} & \underline{5.54} & \underline{20.69} & \underline{2.67} & \underline{1.84} \\
    Only Synthesis & 1.93 & 5.04 & 37.96 & 3.80 & 5.17 \\
    Syn. + Real (Ours) & \underline{1.30} & \textbf{5.93} & \textbf{20.48} & \textbf{2.48} & \textbf{1.82} \\
    \bottomrule
    \end{tabular}
\end{table}

\begin{table}[h!]
    \centering
    \small
    {
    \setlength{\tabcolsep}{2pt}
    \renewcommand{\arraystretch}{1.05}
    \caption{
        Ablation study on conditioning injection strategies.
    }
    \label{tab:ablation_input_style}
    \begin{tabular}{l l cccc}
    \toprule
    Context & Cat. & T2A$\downarrow$ & V2A$\downarrow$ & TTS$\downarrow$ & AE$\downarrow$ \\
    \midrule
    \texttt{mm, trans, sync, mel} & \textit{none} & 40.91 & 25.61 & 32.46 & 5.47 \\
    \texttt{mm, trans, sync} & \texttt{mel} & 38.26 & 26.46 & 25.88 & 4.03 \\
    \textbf{\texttt{mm, trans}} & \textbf{\texttt{sync, mel}} & \textbf{28.90} & \textbf{18.55} & \textbf{19.33} & \textbf{3.62} \\
    \texttt{mm} & \texttt{trans, sync, mel} & 60.58 & 56.88 & 59.20 & 4.88 \\
    \bottomrule
    \end{tabular}
    }
\end{table}

\vspace{1mm}
\noindent\textbf{Ablation on Conditioning Strategies.}
We perform an ablation study to identify the optimal method for integrating our multimodal conditions. We compare four distinct injection strategies by varying how features including multimodal (mm), transcript (trans), synchronization (sync), and mel-spectrogram (mel) are delivered to the DiT backbone, either via cross-attention (Context) or channel-wise concatenation (Cat). As shown in Table~\ref{tab:ablation_input_style}, the results consistently demonstrate the superiority of one particular configuration across all tasks. This optimal strategy involves providing high-level conditions (mm, trans) as flexible Context, while simultaneously concatenating low-level conditions (sync, mel) with the input noise. This result provides the insight for future research on the optimal conditioning strategy for unified audio generation.

\begin{table}[!ht]
    \centering
    \small
    \caption{
        Ablation study on the source of conditional features from the MLLM.
    }
    \label{tab:ablation_feature_source}
    \renewcommand{\arraystretch}{0.9}
    \resizebox{0.43\textwidth}{!}{
    \begin{tabular}{l cccc} 
    \toprule
    \multirow{2}{*}{Feature Source} & \multicolumn{2}{c}{\textbf{T2A}} & \multicolumn{2}{c}{\textbf{T2M}} \\
    \cmidrule(lr){2-3} \cmidrule(lr){4-5}
    & IS$\uparrow$ & FD$\downarrow$ & IS$\uparrow$ & FAD$\downarrow$ \\
    \midrule
    Last Layer (\texttt{-1}) & 9.36 & 4.21 & 2.90 & 3.26 \\
    Penultimate Layer (\texttt{-2}) & \textbf{11.26} & \textbf{2.75} & \textbf{3.46} & \textbf{3.05} \\
    MetaQuery & 7.44 & 8.34 & 2.38 & 8.15 \\
    Query & 8.55 & 5.31 & 2.69 & 4.22 \\
    \bottomrule
    \end{tabular}
    }
\end{table}
\noindent\textbf{Ablation on Feature Source Selection.}
We ablate four strategies for extracting conditional features from the frozen MLLM, with results on T2A and T2M tasks shown in Table~\ref{tab:ablation_feature_source}.
The methods include using the Last Layer (-1) embeddings, the Penultimate Layer (-2) embeddings, MetaQuery (which appends learnable tokens to the input sequence, following~\cite{pan2025transfer}), and Query mechanism (which uses learnable tokens to attend to the penultimate layer features via cross-attention).
Our results indicate that for audio generative tasks, using the unfiltered feature sequence from the penultimate layer is the most effective conditioning strategy. The penultimate layer's superiority confirms that the final layer is overly specialized for text prediction, whereas the penultimate layer retains richer, uncompressed semantic and acoustic details. Furthermore, complex query mechanisms proved detrimental, suggesting that high-fidelity audio synthesis is sensitive to information bottlenecks and benefits most from direct access to dense features.

\section{Conclusion}
\label{conclusion}
In this work, we introduced \model, the first end-to-end framework to unify audio understanding, generation, and editing across the full spectrum of sound, music, and speech. Our decoupled architecture successfully synergizes a frozen MLLM for high-level reasoning with a trainable DiT, guided by a hybrid conditioning mechanism that separates high-level semantic and low-level signal features. To address a critical data bottleneck, we also constructed \editset, a large-scale dataset of over one million instruction-guided editing pairs. Extensive experiments demonstrate that our single unified model not only matches or surpasses the performance of specialized expert models but also exhibits remarkable inherited abilities, including knowledge-augmented reasoning and zero-shot cross-lingual control, inherited from the MLLM. We believe \model provides a powerful and scalable baseline that highlights a promising path toward universal generative audio intelligence.

\clearpage
\bibliographystyle{ACM-Reference-Format}
\bibliography{main}

@String(ICASSP=	{ICASSP})

@String(ICLR = {Int. Conf. Learn. Represent.})

@String(AAAI = {AAAI})

@String(ICLR  = {ICLR})

@article{liu2022flow,
  title={Flow straight and fast: Learning to generate and transfer data with rectified flow},
  author={Liu, Xingchao and Gong, Chengyue and Liu, Qiang},
  journal={arXiv preprint arXiv:2209.03003},
  year={2022}
}

@inproceedings{xing2024seeing,
  title={Seeing and hearing: Open-domain visual-audio generation with diffusion latent aligners},
  author={Xing, Yazhou and He, Yingqing and Tian, Zeyue and Wang, Xintao and Chen, Qifeng},
  booktitle={Proceedings of the IEEE/CVF Conference on Computer Vision and Pattern Recognition},
  pages={7151--7161},
  year={2024}
}

@article{liu2024tell,
  title={Tell What You Hear From What You See--Video to Audio Generation Through Text},
  author={Liu, Xiulong and Su, Kun and Shlizerman, Eli},
  journal={arXiv preprint arXiv:2411.05679},
  year={2024}
}

@article{su2023v2meow,
  title={V2Meow: Meowing to the Visual Beat via Music Generation},
  author={Su, Kun and Li, Judith Yue and Huang, Qingqing and Kuzmin, Dima and Lee, Joonseok and Donahue, Chris and Sha, Fei and Jansen, Aren and Wang, Yu and Verzetti, Mauro and others},
  journal={arXiv preprint arXiv:2305.06594},
  year={2023}
}

@article{lin2024vmas,
  title={VMAS: Video-to-Music Generation via Semantic Alignment in Web Music Videos},
  author={Lin, Yan-Bo and Tian, Yu and Yang, Linjie and Bertasius, Gedas and Wang, Heng},
  journal={arXiv preprint arXiv:2409.07450},
  year={2024}
}

@article{tong2022videomae,
  title={Videomae: Masked autoencoders are data-efficient learners for self-supervised video pre-training},
  author={Tong, Zhan and Song, Yibing and Wang, Jue and Wang, Limin},
  journal={Advances in neural information processing systems},
  volume={35},
  pages={10078--10093},
  year={2022}
}

@inproceedings{kim2019audiocaps,
  title={Audiocaps: Generating captions for audios in the wild},
  author={Kim, Chris Dongjoo and Kim, Byeongchang and Lee, Hyunmin and Kim, Gunhee},
  booktitle={Proceedings of the 2019 Conference of the North American Chapter of the Association for Computational Linguistics: Human Language Technologies, Volume 1 (Long and Short Papers)},
  pages={119--132},
  year={2019}
}

@article{mei2024wavcaps,
  title={Wavcaps: A chatgpt-assisted weakly-labelled audio captioning dataset for audio-language multimodal research},
  author={Mei, Xinhao and Meng, Chutong and Liu, Haohe and Kong, Qiuqiang and Ko, Tom and Zhao, Chengqi and Plumbley, Mark D and Zou, Yuexian and Wang, Wenwu},
  journal={IEEE/ACM Transactions on Audio, Speech, and Language Processing},
  year={2024},
  publisher={IEEE}
}

@inproceedings{chen2020vggsound,
  title={Vggsound: A large-scale audio-visual dataset},
  author={Chen, Honglie and Xie, Weidi and Vedaldi, Andrea and Zisserman, Andrew},
  booktitle={ICASSP 2020-2020 IEEE International Conference on Acoustics, Speech and Signal Processing (ICASSP)},
  pages={721--725},
  year={2020},
  organization={IEEE}
}

@inproceedings{hershey2021benefit,
  title={The benefit of temporally-strong labels in audio event classification},
  author={Hershey, Shawn and Ellis, Daniel PW and Fonseca, Eduardo and Jansen, Aren and Liu, Caroline and Moore, R Channing and Plakal, Manoj},
  booktitle={ICASSP 2021-2021 IEEE International Conference on Acoustics, Speech and Signal Processing (ICASSP)},
  pages={366--370},
  year={2021},
  organization={IEEE}
}

@article{chu2024qwen2,
  title={Qwen2-audio technical report},
  author={Chu, Yunfei and Xu, Jin and Yang, Qian and Wei, Haojie and Wei, Xipin and Guo, Zhifang and Leng, Yichong and Lv, Yuanjun and He, Jinzheng and Lin, Junyang and others},
  journal={arXiv preprint arXiv:2407.10759},
  year={2024}
}

@inproceedings{tian2025vidmuse,
  title={Vidmuse: A simple video-to-music generation framework with long-short-term modeling},
  author={Tian, Zeyue and Liu, Zhaoyang and Yuan, Ruibin and Pan, Jiahao and Liu, Qifeng and Tan, Xu and Chen, Qifeng and Xue, Wei and Guo, Yike},
  booktitle={Proceedings of the Computer Vision and Pattern Recognition Conference},
  pages={18782--18793},
  year={2025}
}

@article{liu2024mumu,
  title={MuMu-LLaMA: Multi-modal Music Understanding and Generation via Large Language Models},
  author={Liu, Shansong and Hussain, Atin Sakkeer and Wu, Qilong and Sun, Chenshuo and Shan, Ying},
  journal={arXiv preprint arXiv:2412.06660},
  year={2024}
}

@article{copet2024simple,
  title={Simple and controllable music generation},
  author={Copet, Jade and Kreuk, Felix and Gat, Itai and Remez, Tal and Kant, David and Synnaeve, Gabriel and Adi, Yossi and D{\'e}fossez, Alexandre},
  journal={Advances in Neural Information Processing Systems},
  volume={36},
  year={2024}
}

@inproceedings{majumder2024tango,
  title={Tango 2: Aligning diffusion-based text-to-audio generations through direct preference optimization},
  author={Majumder, Navonil and Hung, Chia-Yu and Ghosal, Deepanway and Hsu, Wei-Ning and Mihalcea, Rada and Poria, Soujanya},
  booktitle={Proceedings of the 32nd ACM International Conference on Multimedia},
  pages={564--572},
  year={2024}
}

@article{evans2024stable,
  title={Stable audio open},
  author={Evans, Zach and Parker, Julian D and Carr, CJ and Zukowski, Zack and Taylor, Josiah and Pons, Jordi},
  journal={arXiv preprint arXiv:2407.14358},
  year={2024}
}

@article{ghosal2023text,
  title={Text-to-audio generation using instruction-tuned llm and latent diffusion model},
  author={Ghosal, Deepanway and Majumder, Navonil and Mehrish, Ambuj and Poria, Soujanya},
  journal={arXiv preprint arXiv:2304.13731},
  year={2023}
}

@article{kreuk2022audiogen,
  title={Audiogen: Textually guided audio generation},
  author={Kreuk, Felix and Synnaeve, Gabriel and Polyak, Adam and Singer, Uriel and D{\'e}fossez, Alexandre and Copet, Jade and Parikh, Devi and Taigman, Yaniv and Adi, Yossi},
  journal={arXiv preprint arXiv:2209.15352},
  year={2022}
}

@article{kong2020panns,
  title={Panns: Large-scale pretrained audio neural networks for audio pattern recognition},
  author={Kong, Qiuqiang and Cao, Yin and Iqbal, Turab and Wang, Yuxuan and Wang, Wenwu and Plumbley, Mark D},
  journal={IEEE/ACM Transactions on Audio, Speech, and Language Processing},
  volume={28},
  pages={2880--2894},
  year={2020},
  publisher={IEEE}
}

@inproceedings{cheng2025mmaudio,
  title={MMAudio: Taming Multimodal Joint Training for High-Quality Video-to-Audio Synthesis},
  author={Cheng, Ho Kei and Ishii, Masato and Hayakawa, Akio and Shibuya, Takashi and Schwing, Alexander and Mitsufuji, Yuki},
  booktitle={Proceedings of the Computer Vision and Pattern Recognition Conference},
  pages={28901--28911},
  year={2025}
}

@article{he2024llms,
  title={Llms meet multimodal generation and editing: A survey},
  author={He, Yingqing and Liu, Zhaoyang and Chen, Jingye and Tian, Zeyue and Liu, Hongyu and Chi, Xiaowei and Liu, Runtao and Yuan, Ruibin and Xing, Yazhou and Wang, Wenhai and others},
  journal={arXiv preprint arXiv:2405.19334},
  year={2024}
}

@article{deng2024composerx,
  title={Composerx: Multi-agent symbolic music composition with llms},
  author={Deng, Qixin and Yang, Qikai and Yuan, Ruibin and Huang, Yipeng and Wang, Yi and Liu, Xubo and Tian, Zeyue and Pan, Jiahao and Zhang, Ge and Lin, Hanfeng and others},
  journal={arXiv preprint arXiv:2404.18081},
  year={2024}
}

@article{yuan2024chatmusician,
  title={Chatmusician: Understanding and generating music intrinsically with llm},
  author={Yuan, Ruibin and Lin, Hanfeng and Wang, Yi and Tian, Zeyue and Wu, Shangda and Shen, Tianhao and Zhang, Ge and Wu, Yuhang and Liu, Cong and Zhou, Ziya and others},
  journal={arXiv preprint arXiv:2402.16153},
  year={2024}
}

@article{yinghao2024foundation,
  title={Foundation models for music: A survey},
  author={Yinghao, M and Anders, {\O} and Anton, R and Bleiz MacSen Del, S and Charalampos, S and Chris, D},
  journal={arXiv preprint arXiv:2408.14340},
  year={2024}
}

@article{huang2023make,
  title={Make-an-audio 2: Temporal-enhanced text-to-audio generation},
  author={Huang, Jiawei and Ren, Yi and Huang, Rongjie and Yang, Dongchao and Ye, Zhenhui and Zhang, Chen and Liu, Jinglin and Yin, Xiang and Ma, Zejun and Zhao, Zhou},
  journal={arXiv preprint arXiv:2305.18474},
  year={2023}
}

@inproceedings{iashin2024synchformer,
  title={Synchformer: Efficient synchronization from sparse cues},
  author={Iashin, Vladimir and Xie, Weidi and Rahtu, Esa and Zisserman, Andrew},
  booktitle={ICASSP 2024-2024 IEEE International Conference on Acoustics, Speech and Signal Processing (ICASSP)},
  pages={5325--5329},
  year={2024},
  organization={IEEE}
}

@inproceedings{elizalde2023clap,
  title={Clap learning audio concepts from natural language supervision},
  author={Elizalde, Benjamin and Deshmukh, Soham and Al Ismail, Mahmoud and Wang, Huaming},
  booktitle={ICASSP 2023-2023 IEEE International Conference on Acoustics, Speech and Signal Processing (ICASSP)},
  pages={1--5},
  year={2023},
  organization={IEEE}
}

@article{xu2025qwen2,
  title={Qwen2. 5-omni technical report},
  author={Xu, Jin and Guo, Zhifang and He, Jinzheng and Hu, Hangrui and He, Ting and Bai, Shuai and Chen, Keqin and Wang, Jialin and Fan, Yang and Dang, Kai and others},
  journal={arXiv preprint arXiv:2503.20215},
  year={2025}
}

@article{liu2025javisgpt,
  title={Javisgpt: A unified multi-modal llm for sounding-video comprehension and generation},
  author={Liu, Kai and Li, Jungang and Sun, Yuchong and Wu, Shengqiong and Gao, Jianzhang and Zhang, Daoan and Zhang, Wei and Jin, Sheng and Yu, Sicheng and Zhan, Geng and others},
  journal={arXiv preprint arXiv:2512.22905},
  year={2025}
}

@inproceedings{huang2024audiogpt,
  title={Audiogpt: Understanding and generating speech, music, sound, and talking head},
  author={Huang, Rongjie and Li, Mingze and Yang, Dongchao and Shi, Jiatong and Chang, Xuankai and Ye, Zhenhui and Wu, Yuning and Hong, Zhiqing and Huang, Jiawei and Liu, Jinglin and others},
  booktitle={Proceedings of the AAAI Conference on Artificial Intelligence},
  volume={38},
  number={21},
  pages={23802--23804},
  year={2024}
}

@inproceedings{DBLP:conf/aaai/HuangLYSCYWHHLR24,
  author       = {Rongjie Huang and
                  Mingze Li and
                  Dongchao Yang and
                  Jiatong Shi and
                  Xuankai Chang and
                  Zhenhui Ye and
                  Yuning Wu and
                  Zhiqing Hong and
                  Jiawei Huang and
                  Jinglin Liu and
                  Yi Ren and
                  Yuexian Zou and
                  Zhou Zhao and
                  Shinji Watanabe},
  editor       = {Michael J. Wooldridge and
                  Jennifer G. Dy and
                  Sriraam Natarajan},
  title        = {AudioGPT: Understanding and Generating Speech, Music, Sound, and Talking
                  Head},
  booktitle    = {Thirty-Eighth {AAAI} Conference on Artificial Intelligence, {AAAI}
                  2024, Thirty-Sixth Conference on Innovative Applications of Artificial
                  Intelligence, {IAAI} 2024, Fourteenth Symposium on Educational Advances
                  in Artificial Intelligence, {EAAI} 2014, February 20-27, 2024, Vancouver,
                  Canada},
  pages        = {23802--23804},
  publisher    = {{AAAI} Press},
  year         = {2024},
  url          = {https://doi.org/10.1609/aaai.v38i21.30570},
  doi          = {10.1609/AAAI.V38I21.30570},
  bibsource    = {dblp computer science bibliography, https://dblp.org}
}

@article{tian2025ualm,
  title={Ualm: Unified audio language model for understanding, generation and reasoning},
  author={Tian, Jinchuan and Lee, Sang-gil and Kong, Zhifeng and Ghosh, Sreyan and Goel, Arushi and Yang, Chao-Han Huck and Dai, Wenliang and Liu, Zihan and Ye, Hanrong and Watanabe, Shinji and others},
  journal={arXiv preprint arXiv:2510.12000},
  year={2025}
}

@article{wang2023audit,
  title={Audit: Audio editing by following instructions with latent diffusion models},
  author={Wang, Yuancheng and Ju, Zeqian and Tan, Xu and He, Lei and Wu, Zhizheng and Bian, Jiang and others},
  journal={Advances in Neural Information Processing Systems},
  volume={36},
  pages={71340--71357},
  year={2023}
}

@article{tao2025mmedit,
  title={MMEDIT: A Unified Framework for Multi-Type Audio Editing via Audio Language Model},
  author={Tao, Ye and Xu, Xuenan and Wu, Wen and Wang, Shuai and Wu, Mengyue and Zhang, Chao},
  journal={arXiv preprint arXiv:2512.20339},
  year={2025}
}

@inproceedings{DBLP:conf/icml/KongGBPVC24,
  author       = {Zhifeng Kong and
                  Arushi Goel and
                  Rohan Badlani and
                  Wei Ping and
                  Rafael Valle and
                  Bryan Catanzaro},
  title        = {Audio Flamingo: {A} Novel Audio Language Model with Few-Shot Learning
                  and Dialogue Abilities},
  booktitle    = {Forty-first International Conference on Machine Learning, {ICML} 2024,
                  Vienna, Austria, July 21-27, 2024},
  publisher    = {OpenReview.net},
  year         = {2024},
  url          = {https://openreview.net/forum?id=WYi3WKZjYe},
  bibsource    = {dblp computer science bibliography, https://dblp.org}
}

@inproceedings{DBLP:conf/icml/GhoshKKSKPVMC25,
  author       = {Sreyan Ghosh and
                  Zhifeng Kong and
                  Sonal Kumar and
                  S. Sakshi and
                  Jaehyeon Kim and
                  Wei Ping and
                  Rafael Valle and
                  Dinesh Manocha and
                  Bryan Catanzaro},
  title        = {Audio Flamingo 2: An Audio-Language Model with Long-Audio Understanding
                  and Expert Reasoning Abilities},
  booktitle    = {Forty-second International Conference on Machine Learning, {ICML}
                  2025, Vancouver, BC, Canada, July 13-19, 2025},
  publisher    = {OpenReview.net},
  year         = {2025},
  url          = {https://openreview.net/forum?id=xWu5qpDK6U},
  bibsource    = {dblp computer science bibliography, https://dblp.org}
}

@article{chu2023qwen,
  title={Qwen-audio: Advancing universal audio understanding via unified large-scale audio-language models},
  author={Chu, Yunfei and Xu, Jin and Zhou, Xiaohuan and Yang, Qian and Zhang, Shiliang and Yan, Zhijie and Zhou, Chang and Zhou, Jingren},
  journal={arXiv preprint arXiv:2311.07919},
  year={2023}
}

@inproceedings{DBLP:conf/iclr/TangYSC000M024,
  author       = {Changli Tang and
                  Wenyi Yu and
                  Guangzhi Sun and
                  Xianzhao Chen and
                  Tian Tan and
                  Wei Li and
                  Lu Lu and
                  Zejun Ma and
                  Chao Zhang},
  title        = {{SALMONN:} Towards Generic Hearing Abilities for Large Language Models},
  booktitle    = {The Twelfth International Conference on Learning Representations,
                  {ICLR} 2024, Vienna, Austria, May 7-11, 2024},
  publisher    = {OpenReview.net},
  year         = {2024},
  url          = {https://openreview.net/forum?id=14rn7HpKVk},
  bibsource    = {dblp computer science bibliography, https://dblp.org}
}

@article{liu2025thinksound,
  title={Thinksound: Chain-of-thought reasoning in multimodal large language models for audio generation and editing},
  author={Liu, Huadai and Luo, Kaicheng and Wang, Jialei and Wang, Wen and Chen, Qian and Zhao, Zhou and Xue, Wei},
  journal={arXiv preprint arXiv:2506.21448},
  year={2025}
}

@inproceedings{melechovsky2024mustango,
  title={Mustango: Toward controllable text-to-music generation},
  author={Melechovsky, Jan and Guo, Zixun and Ghosal, Deepanway and Majumder, Navonil and Herremans, Dorien and Poria, Soujanya},
  booktitle={Proceedings of the 2024 Conference of the North American Chapter of the Association for Computational Linguistics: Human Language Technologies (Volume 1: Long Papers)},
  pages={8293--8316},
  year={2024}
}

@inproceedings{chen2024musicldm,
  title={Musicldm: Enhancing novelty in text-to-music generation using beat-synchronous mixup strategies},
  author={Chen, Ke and Wu, Yusong and Liu, Haohe and Nezhurina, Marianna and Berg-Kirkpatrick, Taylor and Dubnov, Shlomo},
  booktitle={ICASSP 2024-2024 IEEE International Conference on Acoustics, Speech and Signal Processing (ICASSP)},
  pages={1206--1210},
  year={2024},
  organization={IEEE}
}

@inproceedings{xie2025filmcomposer,
  title={FilmComposer: LLM-Driven Music Production for Silent Film Clips},
  author={Xie, Zhifeng and He, Qile and Zhu, Youjia and He, Qiwei and Li, Mengtian},
  booktitle={Proceedings of the Computer Vision and Pattern Recognition Conference},
  pages={13519--13528},
  year={2025}
}

@article{du2024cosyvoice,
  title={Cosyvoice: A scalable multilingual zero-shot text-to-speech synthesizer based on supervised semantic tokens},
  author={Du, Zhihao and Chen, Qian and Zhang, Shiliang and Hu, Kai and Lu, Heng and Yang, Yexin and Hu, Hangrui and Zheng, Siqi and Gu, Yue and Ma, Ziyang and others},
  journal={arXiv preprint arXiv:2407.05407},
  year={2024}
}

@inproceedings{chen2025f5,
  title={F5-tts: A fairytaler that fakes fluent and faithful speech with flow matching},
  author={Chen, Yushen and Niu, Zhikang and Ma, Ziyang and Deng, Keqi and Wang, Chunhui and JianZhao, JianZhao and Yu, Kai and Chen, Xie},
  booktitle={Proceedings of the 63rd Annual Meeting of the Association for Computational Linguistics (Volume 1: Long Papers)},
  pages={6255--6271},
  year={2025}
}

@article{deng2025indextts,
  title={Indextts: An industrial-level controllable and efficient zero-shot text-to-speech system},
  author={Deng, Wei and Zhou, Siyi and Shu, Jingchen and Wang, Jinchao and Wang, Lu},
  journal={arXiv preprint arXiv:2502.05512},
  year={2025}
}

@article{wang2023neural,
  title={Neural codec language models are zero-shot text to speech synthesizers},
  author={Wang, Chengyi and Chen, Sanyuan and Wu, Yu and Zhang, Ziqiang and Zhou, Long and Liu, Shujie and Chen, Zhuo and Liu, Yanqing and Wang, Huaming and Li, Jinyu and others},
  journal={arXiv preprint arXiv:2301.02111},
  year={2023}
}

@article{peng2024voicecraft,
  title={Voicecraft: Zero-shot speech editing and text-to-speech in the wild},
  author={Peng, Puyuan and Huang, Po-Yao and Li, Shang-Wen and Mohamed, Abdelrahman and Harwath, David},
  journal={arXiv preprint arXiv:2403.16973},
  year={2024}
}

@article{tian2025audiox,
  title={Audiox: Diffusion transformer for anything-to-audio generation},
  author={Tian, Zeyue and Jin, Yizhu and Liu, Zhaoyang and Yuan, Ruibin and Tan, Xu and Chen, Qifeng and Xue, Wei and Guo, Yike},
  journal={arXiv preprint arXiv:2503.10522},
  year={2025}
}

@article{wang2025audiogen,
  title={Audiogen-omni: A unified multimodal diffusion transformer for video-synchronized audio, speech, and song generation},
  author={Wang, Le and Wang, Jun and Qiang, Chunyu and Deng, Feng and Zhang, Chen and Zhang, Di and Gai, Kun},
  journal={arXiv preprint arXiv:2508.00733},
  year={2025}
}

@article{liu2025unimoe,
  title={UniMoE-Audio: Unified Speech and Music Generation with Dynamic-Capacity MoE},
  author={Liu, Zhenyu and Li, Yunxin and Zhang, Xuanyu and Teng, Qixun and Jiang, Shenyuan and Chen, Xinyu and Shi, Haoyuan and Li, Jinchao and Wang, Qi and Chen, Haolan and others},
  journal={arXiv preprint arXiv:2510.13344},
  year={2025}
}

@inproceedings{rong2025audiogenie,
  title={Audiogenie: A training-free multi-agent framework for diverse multimodality-to-multiaudio generation},
  author={Rong, Yan and Wang, Jinting and Lei, Guangzhi and Yang, Shan and Liu, Li},
  booktitle={Proceedings of the 33rd ACM International Conference on Multimedia},
  pages={8872--8881},
  year={2025}
}

@article{manor2024zero,
  title={Zero-shot unsupervised and text-based audio editing using DDPM inversion},
  author={Manor, Hila and Michaeli, Tomer},
  journal={arXiv preprint arXiv:2402.10009},
  year={2024}
}

@article{lan2025guiding,
  title={Guiding audio editing with audio language model},
  author={Lan, Zitong and Hao, Yiduo and Zhao, Mingmin},
  journal={arXiv preprint arXiv:2509.21625},
  year={2025}
}

@article{pan2025transfer,
  title={Transfer between modalities with metaqueries},
  author={Pan, Xichen and Shukla, Satya Narayan and Singh, Aashu and Zhao, Zhuokai and Mishra, Shlok Kumar and Wang, Jialiang and Xu, Zhiyang and Chen, Jiuhai and Li, Kunpeng and Juefei-Xu, Felix and others},
  journal={arXiv preprint arXiv:2504.06256},
  year={2025}
}

@article{jiao2025thinkgen,
  title={ThinkGen: Generalized Thinking for Visual Generation},
  author={Jiao, Siyu and Lin, Yiheng and Zhong, Yujie and She, Qi and Zhou, Wei and Lan, Xiaohan and Huang, Zilong and Yu, Fei and Yu, Yingchen and Zhao, Yunqing and others},
  journal={arXiv preprint arXiv:2512.23568},
  year={2025}
}

@inproceedings{ma2025janusflow,
  title={Janusflow: Harmonizing autoregression and rectified flow for unified multimodal understanding and generation},
  author={Ma, Yiyang and Liu, Xingchao and Chen, Xiaokang and Liu, Wen and Wu, Chengyue and Wu, Zhiyu and Pan, Zizheng and Xie, Zhenda and Zhang, Haowei and Yu, Xingkai and others},
  booktitle={Proceedings of the Computer Vision and Pattern Recognition Conference},
  pages={7739--7751},
  year={2025}
}

@article{chen2025janus,
  title={Janus-pro: Unified multimodal understanding and generation with data and model scaling},
  author={Chen, Xiaokang and Wu, Zhiyu and Liu, Xingchao and Pan, Zizheng and Liu, Wen and Xie, Zhenda and Yu, Xingkai and Ruan, Chong},
  journal={arXiv preprint arXiv:2501.17811},
  year={2025}
}

@article{wei2025univideo,
  title={Univideo: Unified understanding, generation, and editing for videos},
  author={Wei, Cong and Liu, Quande and Ye, Zixuan and Wang, Qiulin and Wang, Xintao and Wan, Pengfei and Gai, Kun and Chen, Wenhu},
  journal={arXiv preprint arXiv:2510.08377},
  year={2025}
}

@inproceedings{zhan2024anygpt,
  title={Anygpt: Unified multimodal llm with discrete sequence modeling},
  author={Zhan, Jun and Dai, Junqi and Ye, Jiasheng and Zhou, Yunhua and Zhang, Dong and Liu, Zhigeng and Zhang, Xin and Yuan, Ruibin and Zhang, Ge and Li, Linyang and others},
  booktitle={Proceedings of the 62nd Annual Meeting of the Association for Computational Linguistics (Volume 1: Long Papers)},
  pages={9637--9662},
  year={2024}
}

@inproceedings{wu2024next,
  title={Next-gpt: Any-to-any multimodal llm},
  author={Wu, Shengqiong and Fei, Hao and Qu, Leigang and Ji, Wei and Chua, Tat-Seng},
  booktitle={Forty-first International Conference on Machine Learning},
  year={2024}
}

@inproceedings{lu2024unified,
  title={Unified-io 2: Scaling autoregressive multimodal models with vision language audio and action},
  author={Lu, Jiasen and Clark, Christopher and Lee, Sangho and Zhang, Zichen and Khosla, Savya and Marten, Ryan and Hoiem, Derek and Kembhavi, Aniruddha},
  booktitle={Proceedings of the IEEE/CVF Conference on Computer Vision and Pattern Recognition},
  pages={26439--26455},
  year={2024}
}

@article{an2024funaudiollm,
  title={Funaudiollm: Voice understanding and generation foundation models for natural interaction between humans and llms},
  author={An, Keyu and Chen, Qian and Deng, Chong and Du, Zhihao and Gao, Changfeng and Gao, Zhifu and Gu, Yue and He, Ting and Hu, Hangrui and Hu, Kai and others},
  journal={arXiv preprint arXiv:2407.04051},
  year={2024}
}

@article{ai2025ming,
  title={Ming-Omni: A Unified Multimodal Model for Perception and Generation},
  author={AI, Inclusion and Gong, Biao and Zou, Cheng and Zheng, Chuanyang and Zhou, Chunluan and Yan, Canxiang and Jin, Chunxiang and Shen, Chunjie and Zheng, Dandan and Wang, Fudong and others},
  journal={arXiv preprint arXiv:2506.09344},
  year={2025}
}

@article{shi2025sam,
  title={SAM Audio: Segment Anything in Audio},
  author={Shi, Bowen and Tjandra, Andros and Hoffman, John and Wang, Helin and Wu, Yi-Chiao and Gao, Luya and Richter, Julius and Le, Matt and Vyas, Apoorv and Chen, Sanyuan and others},
  journal={arXiv preprint arXiv:2512.18099},
  year={2025}
}

@inproceedings{salamon2017scaper,
  title={Scaper: A library for soundscape synthesis and augmentation},
  author={Salamon, Justin and MacConnell, Duncan and Cartwright, Mark and Li, Peter and Bello, Juan Pablo},
  booktitle={2017 IEEE Workshop on Applications of Signal Processing to Audio and Acoustics (WASPAA)},
  pages={344--348},
  year={2017},
  organization={IEEE}
}

@inproceedings{piczak2015esc,
  title={ESC: Dataset for environmental sound classification},
  author={Piczak, Karol J},
  booktitle={Proceedings of the 23rd ACM international conference on Multimedia},
  pages={1015--1018},
  year={2015}
}

@article{bai2025audiosetcaps,
  title={Audiosetcaps: An enriched audio-caption dataset using automated generation pipeline with large audio and language models},
  author={Bai, Jisheng and Liu, Haohe and Wang, Mou and Shi, Dongyuan and Wang, Wenwu and Plumbley, Mark D and Gan, Woon-Seng and Chen, Jianfeng},
  journal={IEEE Transactions on Audio, Speech and Language Processing},
  year={2025},
  publisher={IEEE}
}

@article{xue2025audio,
  title={Audio-flan: A preliminary release},
  author={Xue, Liumeng and Zhou, Ziya and Pan, Jiahao and Li, Zixuan and Fan, Shuai and Ma, Yinghao and Cheng, Sitong and Yang, Dongchao and Guo, Haohan and Xiao, Yujia and others},
  journal={arXiv preprint arXiv:2502.16584},
  year={2025}
}

@article{wang2025mmsu,
  title={MMSU: A Massive Multi-task Spoken Language Understanding and Reasoning Benchmark},
  author={Wang, Dingdong and Wu, Jincenzi and Li, Junan and Yang, Dongchao and Chen, Xueyuan and Zhang, Tianhua and Meng, Helen},
  journal={arXiv preprint arXiv:2506.04779},
  year={2025}
}

@article{sakshi2024mmau,
  title={Mmau: A massive multi-task audio understanding and reasoning benchmark},
  author={Sakshi, S and Tyagi, Utkarsh and Kumar, Sonal and Seth, Ashish and Selvakumar, Ramaneswaran and Nieto, Oriol and Duraiswami, Ramani and Ghosh, Sreyan and Manocha, Dinesh},
  journal={arXiv preprint arXiv:2410.19168},
  year={2024}
}

@article{raffel2020exploring,
  title={Exploring the limits of transfer learning with a unified text-to-text transformer},
  author={Raffel, Colin and Shazeer, Noam and Roberts, Adam and Lee, Katherine and Narang, Sharan and Matena, Michael and Zhou, Yanqi and Li, Wei and Liu, Peter J},
  journal={Journal of machine learning research},
  volume={21},
  number={140},
  pages={1--67},
  year={2020}
}

@inproceedings{radford2021learning,
  title={Learning transferable visual models from natural language supervision},
  author={Radford, Alec and Kim, Jong Wook and Hallacy, Chris and Ramesh, Aditya and Goh, Gabriel and Agarwal, Sandhini and Sastry, Girish and Askell, Amanda and Mishkin, Pamela and Clark, Jack and others},
  booktitle={International conference on machine learning},
  pages={8748--8763},
  year={2021},
  organization={PmLR}
}

@article{goel2025audio,
  title={Audio flamingo 3: Advancing audio intelligence with fully open large audio language models},
  author={Goel, Arushi and Ghosh, Sreyan and Kim, Jaehyeon and Kumar, Sonal and Kong, Zhifeng and Lee, Sang-gil and Yang, Chao-Han Huck and Duraiswami, Ramani and Manocha, Dinesh and Valle, Rafael and others},
  journal={arXiv preprint arXiv:2507.08128},
  year={2025}
}

@article{wang2024maskgct,
  title={Maskgct: Zero-shot text-to-speech with masked generative codec transformer},
  author={Wang, Yuancheng and Zhan, Haoyue and Liu, Liwei and Zeng, Ruihong and Guo, Haotian and Zheng, Jiachen and Zhang, Qiang and Zhang, Xueyao and Zhang, Shunsi and Wu, Zhizheng},
  journal={arXiv preprint arXiv:2409.00750},
  year={2024}
}

@article{du2025cosyvoice,
  title={Cosyvoice 3: Towards in-the-wild speech generation via scaling-up and post-training},
  author={Du, Zhihao and Gao, Changfeng and Wang, Yuxuan and Yu, Fan and Zhao, Tianyu and Wang, Hao and Lv, Xiang and Wang, Hui and Ni, Chongjia and Shi, Xian and others},
  journal={arXiv preprint arXiv:2505.17589},
  year={2025}
}

@article{meng2021sdedit,
  title={Sdedit: Guided image synthesis and editing with stochastic differential equations},
  author={Meng, Chenlin and He, Yutong and Song, Yang and Song, Jiaming and Wu, Jiajun and Zhu, Jun-Yan and Ermon, Stefano},
  journal={arXiv preprint arXiv:2108.01073},
  year={2021}
}

@article{anastassiou2024seed,
  title={Seed-tts: A family of high-quality versatile speech generation models},
  author={Anastassiou, Philip and Chen, Jiawei and Chen, Jitong and Chen, Yuanzhe and Chen, Zhuo and Chen, Ziyi and Cong, Jian and Deng, Lelai and Ding, Chuang and Gao, Lu and others},
  journal={arXiv preprint arXiv:2406.02430},
  year={2024}
}

@inproceedings{xie2025audiotime,
  title={Audiotime: A temporally-aligned audio-text benchmark dataset},
  author={Xie, Zeyu and Xu, Xuenan and Wu, Zhizheng and Wu, Mengyue},
  booktitle={ICASSP 2025-2025 IEEE International Conference on Acoustics, Speech and Signal Processing (ICASSP)},
  pages={1--5},
  year={2025},
  organization={IEEE}
}

@article{yuan2025yue,
  title={Yue: Scaling open foundation models for long-form music generation},
  author={Yuan, Ruibin and Lin, Hanfeng and Guo, Shuyue and Zhang, Ge and Pan, Jiahao and Zang, Yongyi and Liu, Haohe and Liang, Yiming and Ma, Wenye and Du, Xingjian and others},
  journal={arXiv preprint arXiv:2503.08638},
  year={2025}
}

@article{chi2024mmtrail,
  title={Mmtrail: A multimodal trailer video dataset with language and music descriptions},
  author={Chi, Xiaowei and Wang, Yatian and Cheng, Aosong and Fang, Pengjun and Tian, Zeyue and He, Yingqing and Liu, Zhaoyang and Qi, Xingqun and Pan, Jiahao and Zhang, Rongyu and others},
  journal={arXiv preprint arXiv:2407.20962},
  year={2024}
}

\clearpage
\setcounter{table}{0}   
\setcounter{figure}{0}
\setcounter{section}{0}
\renewcommand{\thetable}{A\arabic{table}}
\renewcommand{\thefigure}{A\arabic{figure}}

\section{Appendix}
\label{sec:Appendix}

\begin{table}[!ht]
    \centering
    \small
    \caption{
        \textbf{Ablation study on different encoders.} We evaluate the impact of different encoders for text-to-audio (T2A) and video-to-audio (V2A) tasks.
    }
    \renewcommand{\arraystretch}{1.1}
    \begin{tabular}{ll cccc}
    \toprule
    \textbf{Task} & \textbf{Encoder} & \textbf{KL}$\downarrow$ & \textbf{IS}$\uparrow$ & \textbf{FD}$\downarrow$ & \textbf{FAD}$\downarrow$ \\
    \midrule
    \multirow{3}{*}{T2A} 
    & T5 & 1.51 & 11.27 & 15.78 & 2.75 \\
    & CLAP & 1.81 & 9.64 & 16.58 & \textbf{1.84} \\
    & Qwen & \textbf{1.25} & \textbf{11.65} & \textbf{12.21} & 2.14 \\
    \midrule
    \multirow{3}{*}{V2A}
    & CLIP & 2.20 & 7.42 & 12.21 & 2.80 \\
    & VideoMAE & 2.29 & 6.58 & 14.72 & 3.06 \\
    & Qwen & \textbf{2.00} & \textbf{8.08} & \textbf{8.50} & \textbf{2.23} \\
    \bottomrule
    \end{tabular}
    \label{tab:ablation_encoders}
\end{table}

\begin{table}[!ht]
    \centering
    \scriptsize
    \caption{
        Zero-shot cross-lingual text-to-audio generation results.
    }
    \renewcommand{\arraystretch}{1.05}
    \setlength{\tabcolsep}{6pt}
    \begin{tabular}{l c cccc}
    \toprule
    Method & Language & KL$\downarrow$ & IS$\uparrow$ & FD$\downarrow$ & FAD$\downarrow$ \\
    \midrule
    Tango2~\cite{majumder2024tango} & EN & \textbf{1.11} & 10.37 & 12.22 & 3.20 \\
    AudioX~\cite{tian2025audiox} & EN & 1.34 & \textbf{12.09} & \textbf{11.83} & \textbf{1.86}  \\
    MMAudio~\cite{cheng2025mmaudio} & EN & 1.35 & 12.03 & 12.63 & 4.71 \\
    Stable-Audio-Open~\cite{evans2024stable} & EN & 2.01 & 10.37 & 29.01 & 3.15 \\
    \cmidrule(lr){1-6}
    \multirow{6}{*}{\model} & EN & \underline{1.15} & \underline{11.64} & \underline{11.97} & \textbf{1.86} \\
    & CN & 1.65 & 11.10 & 15.05 & 2.26 \\
    & ES & 2.36 & 9.16 & 25.26 & 4.32 \\
    & DE & 2.39 & 9.13 & 23.51 & 2.92 \\
    & FR & 2.47 & 8.80 & 28.63 & 4.21 \\
    & JP & 2.27 & 8.67 & 19.82 & 3.13 \\
    \bottomrule
    \end{tabular}
    \label{tab:crosslingual_t2a}
\end{table}

\begin{table*}[!htb]
    \centering
    \scriptsize
    \setlength{\tabcolsep}{2pt}
    \caption{
        Detailed results on audio editing tasks. We report FAD/LSD for each task and their average.
    }
    \label{tab:appendix_editing_tasks}
    \renewcommand{\arraystretch}{0.75}
    \resizebox{0.85\textwidth}{!}{
    \begin{tabular}{l cc cc cc cc cc}
    \toprule
    \multirow{2}{*}{Method} &
    \multicolumn{2}{c}{Add} &
    \multicolumn{2}{c}{Remove} &
    \multicolumn{2}{c}{Extract} &
    \multicolumn{2}{c}{Style Transfer} &
    \multicolumn{2}{c}{Avg.} \\
    \cmidrule(lr){2-3} \cmidrule(lr){4-5} \cmidrule(lr){6-7} \cmidrule(lr){8-9} \cmidrule(lr){10-11}
    & FAD$\downarrow$ & LSD$\downarrow$ & FAD$\downarrow$ & LSD$\downarrow$ & FAD$\downarrow$ & LSD$\downarrow$ & FAD$\downarrow$ & LSD$\downarrow$ & FAD$\downarrow$ & LSD$\downarrow$ \\
    \midrule
    ZETA~\cite{manor2024zero} & 5.36 & 4.15 & 3.43 & 4.19 & 2.86 & 3.63 & 3.59 & 3.27 & 3.81 & 3.81 \\
    SDEdit~\cite{meng2021sdedit} & 4.22 & 4.18 & 2.16 & 4.46 & 4.51 & 4.31 & 3.14 & 4.63 & 3.51 & 4.40 \\
    MMEDIT~\cite{tao2025mmedit} & 5.30 & 3.55 & 2.08 & 4.41 & 4.54 & 3.88 & 3.88 & 4.35 & 3.95 & 4.05 \\
    \cellcolor{cyan!7}\model & \cellcolor{cyan!7}2.78 & \cellcolor{cyan!7}2.32 & \cellcolor{cyan!7}3.68 & \cellcolor{cyan!7}1.91 & \cellcolor{cyan!7}3.52 & \cellcolor{cyan!7}2.04 & \cellcolor{cyan!7}3.11 & \cellcolor{cyan!7}2.82 & \cellcolor{cyan!7}3.27 & \cellcolor{cyan!7}2.27 \\
    \bottomrule
    \end{tabular}
    }
\end{table*}

\begin{table}[!ht]
    \centering
    \small
    \caption{
        Training data summary across tasks.
    }
    \label{tab:training_datasets}
    \renewcommand{\arraystretch}{0.9}
    \setlength{\tabcolsep}{4pt}
    \resizebox{\columnwidth}{!}{
    \begin{tabular}{>{\centering\arraybackslash}m{0.12\columnwidth} >{\centering\arraybackslash}m{0.10\columnwidth} >{\arraybackslash}m{0.62\columnwidth}}
    \toprule
    \textbf{Task} & \textbf{Hours} & \textbf{Datasets} \\
    \midrule
    T2A & 1.4k & AudioCaps~\cite{kim2019audiocaps}\newline WavCaps~\cite{mei2024wavcaps}\newline AudioSetCaps~\cite{bai2025audiosetcaps}\newline AudioTime~\cite{xie2025audiotime}\newline IF-Caps~\cite{tian2025audiox} \\
    \midrule
    V2A & 0.7k & VGGSound~\cite{chen2020vggsound}\newline AudioSet Strong~\cite{hershey2021benefit} \\
    \midrule
    T2M & 17k & IF-Caps~\cite{tian2025audiox}\newline MUCaps~\cite{liu2024mumu} \\
    \midrule
    V2M & 16k & V2M~\cite{tian2025vidmuse}\newline MMTrail~\cite{chi2024mmtrail} \\
    \midrule
    Speech & 6k & Audio-FLAN (English subset)~\cite{xue2025audio} \\
    \midrule
    Editing & 3k & \editset (Sec.~\ref{sec:dataset}) \\
    \bottomrule
    \end{tabular}
    }
\end{table}

This section includes dataset details, zero-shot cross-lingual text-to-audio generation, and ablation studies.

\subsection{Dataset Details}
\label{sec:appendix_dataset_details}
Our model is trained on a mixture of diverse datasets, as detailed in Table~\ref{tab:training_datasets}. The data composition for each task is as follows:

\noindent\textbf{Text-to-Audio (T2A):} Approximately 1.4k hours, sourced from AudioCaps~\cite{kim2019audiocaps}, WavCaps~\cite{mei2024wavcaps}, AudioSetCaps~\cite{bai2025audiosetcaps}, and AudioTime~\cite{xie2025audiotime}.

\noindent\textbf{Video-to-Audio (V2A):} Approximately 700 hours, sourced from VGGSound~\cite{chen2020vggsound} and the AudioSet Strong~\cite{hershey2021benefit} benchmark.

\noindent\textbf{Text-to-Music (T2M):} Approximately 17k hours, combining data from V2M~\cite{tian2025vidmuse} and MUCaps~\cite{liu2024mumu}.

\noindent\textbf{Video-to-Music (V2M):} Approximately 16k hours, sourced entirely from the V2M~\cite{tian2025vidmuse} dataset.

\noindent\textbf{Speech:} Approximately 6k hours, using the English subset of Audio-FLAN~\cite{xue2025audio}.

\noindent\textbf{Audio Editing:} Approximately 3k hours from our internally constructed \editset dataset, with the methodology detailed in Section~\ref{sec:dataset}.

\noindent\textbf{Style-Transfer Keyword Generation.}
For style-transfer data construction, we prompt Gemini 2.5 Pro with each audio's original keyword to generate semantically related but stylistically different target keywords. The prompt template is: \textit{``Given audio with keyword `[original\_keyword]', generate a related but different keyword for style transfer.''} Generated candidates are filtered by CLAP before guiding ZETA for style transformation.

\subsection{More Comparison Results}

\noindent\textbf{Detailed Results on Audio Editing Tasks.}
Table~\ref{tab:appendix_editing_tasks} presents detailed performance on the four primary audio editing tasks using FAD and LSD on our \editset benchmark. \model consistently achieves SOTA performance across all individual tasks.

\noindent\textbf{Detailed Results on Generation Tasks.}
We provide a comprehensive breakdown of our model's performance across all generation tasks with multiple evaluation metrics in Table~\ref{tab:appendix_generation_detailed}. The table presents results on KL divergence, Inception Score (IS), Fréchet Distance (FD), and Fréchet Audio Distance (FAD) for Text-to-Audio (T2A), Text-to-Music (T2M), Video-to-Audio (V2A), Video-to-Music (V2M), and Text-to-Speech (TTS) tasks.

\begin{table}[t]
    \centering
    \tiny
    \caption{
        Detailed quantitative results on multimodal generation benchmarks with multiple metrics.
    }
    \renewcommand{\arraystretch}{1.0}
    \resizebox{\columnwidth}{!}{
    \setlength{\tabcolsep}{3pt}
    \begin{tabular}{l cccc}
    \toprule
    Method & KL $\downarrow$ & IS $\uparrow$ & FD $\downarrow$ & FAD $\downarrow$ \\
    \midrule
    \multicolumn{5}{c}{\textit{Text-to-Audio (T2A)}} \\
    \midrule
    Tango2~\cite{majumder2024tango} & \textbf{1.11} & 10.37 & 12.22 & 3.20 \\
    MMAudio~\cite{cheng2025mmaudio} & 1.35 & \underline{12.03} & 12.63 & 4.71 \\
    Stable-Audio-Open~\cite{evans2024stable} & 2.01 & 10.37 & 29.01 & 3.15 \\
    Unified-IO2~\cite{lu2024unified} & 2.72 & 5.44 & 37.95 & 7.81 \\
    AudioX~\cite{tian2025audiox} & 1.34 & \textbf{12.09} & \textbf{11.83} & \textbf{1.86} \\
    \cellcolor{cyan!7} \model & \cellcolor{cyan!7} \underline{1.15} & \cellcolor{cyan!7} 11.64 & \cellcolor{cyan!7} \underline{11.97} & \cellcolor{cyan!7} \textbf{1.86} \\
    \midrule
    \multicolumn{5}{c}{\textit{Text-to-Music (T2M)}} \\
    \midrule
    Stable-Audio-Open~\cite{evans2024stable} & 1.51 & 2.94 & 36.33 & 3.23 \\
    MusicGen~\cite{copet2024simple} & 1.43 & 2.24 & 25.40 & 4.55 \\
    AudioX~\cite{tian2025audiox} & 1.02 & \textbf{3.54} & \underline{10.63} & \underline{1.53} \\
    MuMuLLaMA~\cite{liu2024mumu} & 1.00 & 1.25 & 52.25 & 5.10 \\
    Unified-IO2~\cite{lu2024unified} & \textbf{0.81} & 2.47 & 18.94 & 3.17 \\
    \cellcolor{cyan!7} \model & \cellcolor{cyan!7} \underline{0.84} & \cellcolor{cyan!7} \underline{3.49} & \cellcolor{cyan!7} \textbf{8.68} & \cellcolor{cyan!7} \textbf{1.41} \\
    \midrule
    \multicolumn{5}{c}{\textit{Video-to-Audio (V2A)}} \\
    \midrule
    MMAudio~\cite{cheng2025mmaudio} & \underline{1.97} & \textbf{14.95} & \textbf{6.18} & 2.04 \\
    VATT~\cite{liu2024tell} & \textbf{1.40} & 10.02 & 11.71 & 2.55 \\
    AudioX~\cite{tian2025audiox} & 2.57 & \underline{12.16} & 8.83 & \textbf{1.13} \\
    \cellcolor{cyan!7} \model & \cellcolor{cyan!7} 1.98 & \cellcolor{cyan!7} 10.35 & \cellcolor{cyan!7} \underline{8.33} & \cellcolor{cyan!7} \underline{1.71} \\
    \midrule
    \multicolumn{5}{c}{\textit{Video-to-Music (V2M)}} \\
    \midrule
    VidMuse~\cite{tian2025vidmuse} & 0.73 & 1.32 & \underline{22.95} & 2.46 \\
    AudioX~\cite{tian2025audiox} & \underline{0.69} & \underline{1.34} & 23.96 & \underline{2.12} \\
    MuMuLLaMA~\cite{liu2024mumu} & 1.00 & 1.25 & 52.25 & 5.10 \\
    \cellcolor{cyan!7} \model & \cellcolor{cyan!7} \textbf{0.64} & \cellcolor{cyan!7} \textbf{1.36} & \cellcolor{cyan!7} \textbf{17.46} & \cellcolor{cyan!7} \textbf{1.51} \\
    \bottomrule
    \end{tabular}
    }
    \label{tab:appendix_generation_detailed}
\end{table}

\noindent\textbf{Zero-shot Cross-lingual Text-to-Audio Generation.}
A remarkable inherited capability of our framework is its zero-shot cross-lingual generation, inherited directly from the frozen MLLM's multilingual understanding. To evaluate this, we translated the AudioCaps test set into Chinese (CN), Spanish (ES), German (DE), French (FR), and Japanese (JP) using Gemini. As shown in Table~\ref{tab:crosslingual_t2a}, \model maintains strong performance across all languages despite being trained almost exclusively on English. Notably, the quality of audio generated from non-English prompts (e.g., Chinese) is comparable to strong, English-only specialist models. This result validates that our decoupled architecture effectively transfers the MLLM's linguistic capabilities to the synthesis task, bridging the language gap in generative audio.

\subsection{More Ablation Studies}
\label{sec:appendix_ablation_studies}

\vspace{1mm}

\vspace{1mm}
\noindent\textbf{Effect of the Unified MLLM Encoder.}
To validate our choice of the understanding module, we compare our frozen Qwen-Omni-3B (Qwen) MLLM against specialized single-modality encoders. For T2A, evaluated on AudioCaps~\cite{kim2019audiocaps}, we replace it with text encoders (T5~\cite{raffel2020exploring}, CLAP~\cite{elizalde2023clap}). For V2A, evaluated on VGGSound~\cite{chen2020vggsound}, we compare against vision encoders (CLIP~\cite{radford2021learning}, VideoMAE~\cite{tong2022videomae}). As shown in Table~\ref{tab:ablation_encoders}, the results consistently demonstrate the superiority of using a unified MLLM, which achieves significantly better performance across all metrics in both tasks. We attribute this to the MLLM's richer semantic understanding and, more importantly, its inherent ability to process multimodal contexts jointly, capturing cross-modal relationships that specialized encoders inherently miss.

\subsection{Human Evaluation}
\label{sec:appendix_human_evaluation}

To complement our objective metrics, we perform a comprehensive human evaluation study with 20 audio professionals. Following the evaluation protocols established in prior audio generation works~\cite{tian2025audiox,majumder2024tango,kreuk2022audiogen}, we assess the perceptual quality of our generated outputs against competitive baselines. For each task (T2A, T2M, V2A, V2M, and Audio Editing), we randomly select 20 test samples and present them to raters in a randomized, anonymized manner. Raters are asked to score each sample on two key dimensions: Overall Quality (OVL), which measures the general audio fidelity and naturalness, and Relevance (REL), which evaluates how well the output aligns with the given condition (text prompt, video content, or editing instruction). Both metrics are rated on a scale from 1 to 100, with higher scores indicating better performance.

As shown in Table~\ref{tab:human_evaluation}, \model consistently achieves competitive or superior ratings across all evaluated tasks, demonstrating that our model's outputs align well with human perception of quality and instruction adherence.

\begin{table}[H]
    \centering
    \small
    \caption{
        Human evaluation results on generation and editing tasks.
    }
    \renewcommand{\arraystretch}{0.9}
    \setlength{\tabcolsep}{5pt}
    \begin{tabular}{l l cc}
    \toprule
    Task & Method & OVL$\uparrow$ & REL$\uparrow$ \\
    \midrule
    \multirow{3}{*}{T2A} & Tango2~\cite{majumder2024tango} & 72.3 & 74.1 \\
    & AudioX~\cite{tian2025audiox} & \textbf{81.5} & \underline{83.2} \\
    \cellcolor{cyan!7} & \cellcolor{cyan!7} \model & \cellcolor{cyan!7} \underline{78.6} & \cellcolor{cyan!7} \textbf{83.5} \\
    \midrule
    \multirow{3}{*}{T2M} & MusicGen~\cite{copet2024simple} & 70.8 & 72.5 \\
    & AudioX~\cite{tian2025audiox} & \underline{79.2} & \underline{81.3} \\
    \cellcolor{cyan!7} & \cellcolor{cyan!7} \model & \cellcolor{cyan!7} \textbf{82.7} & \cellcolor{cyan!7} \textbf{81.6} \\
    \midrule
    \multirow{3}{*}{V2A} & MMAudio~\cite{cheng2025mmaudio} & \textbf{80.2} & \textbf{81.8} \\
    & AudioX~\cite{tian2025audiox} & \underline{79.5} & \underline{81.2} \\
    \cellcolor{cyan!7} & \cellcolor{cyan!7} \model & \cellcolor{cyan!7} 75.3 & \cellcolor{cyan!7} 77.1 \\
    \midrule
    \multirow{3}{*}{V2M} & VidMuse~\cite{tian2025vidmuse} & 73.5 & 75.2 \\
    & AudioX~\cite{tian2025audiox} & \underline{78.9} & \underline{80.7} \\
    \cellcolor{cyan!7} & \cellcolor{cyan!7} \model & \cellcolor{cyan!7} \textbf{80.3} & \cellcolor{cyan!7} \textbf{81.0} \\
    \midrule
    \multirow{3}{*}{Editing} & SDEdit~\cite{meng2021sdedit} & 68.4 & 70.2 \\
    & ZETA~\cite{manor2024zero} & \underline{74.6} & \underline{76.3} \\
    \cellcolor{cyan!7} & \cellcolor{cyan!7} \model & \cellcolor{cyan!7} \textbf{79.8} & \cellcolor{cyan!7} \textbf{81.5} \\
    \bottomrule
    \end{tabular}
    \label{tab:human_evaluation}
\end{table}




\section{Ethics Statement}
\label{sec:ethics}

While \model demonstrates powerful capabilities in audio understanding, generation, and editing, we acknowledge potential ethical risks associated with generative audio technologies. Tasks such as voice conversion and speech synthesis could be misused for creating deepfakes, impersonation, or spreading misinformation. To mitigate these risks, we will require users to accept responsible-use terms before accessing the model, explicitly prohibiting malicious applications. We encourage the community to develop robust audio watermarking and detection methods, and believe that with proper safeguards, \model can serve as a valuable tool for creative expression and scientific research.

\end{document}